\documentclass[]{iopart}
\usepackage{graphicx, epsfig} 
\usepackage{bm}

\expandafter\let\csname equation*\endcsname\relax
\expandafter\let\csname endequation*\endcsname\relax

\usepackage[caption=false]{subfig}
\usepackage[usenames]{color}

\usepackage{marvosym}
\usepackage{yfonts}
\usepackage{upgreek}
\usepackage{pdflscape}
\definecolor{linkcolor}{rgb}{0.0,0.3,0.5}
\definecolor{purple}{rgb}{0.7,0.05,0.5}
\usepackage[unicode,colorlinks=true,citecolor=linkcolor,linkcolor=linkcolor,urlcolor=linkcolor]{hyperref}
\usepackage[all]{hypcap}
\usepackage{amsmath,amsfonts,amssymb}
\usepackage{bm}
\usepackage{longtable}
\usepackage[T1]{fontenc}
\usepackage[utf8]{inputenc}
\usepackage{verbatim}
\usepackage{pifont}
\usepackage{yfonts}
\usepackage{url}
\usepackage{multirow}
\usepackage{tensor}
\usepackage{mathtools}
\usepackage[normalem]{ulem}
\usepackage{fancyhdr}

\newcommand{\A}{{\scriptscriptstyle{A}}}
\newcommand{\B}{{\scriptscriptstyle{B}}}
\newcommand{\CC}{{\scriptscriptstyle{C}}}

\newcommand{\I}{{\scriptscriptstyle{I}}}
\newcommand{\J}{{\scriptscriptstyle{J}}}
\newcommand{\M}{{\scriptscriptstyle{M}}}
\newcommand{\N}{{\scriptscriptstyle{N}}}
\newcommand{\K}{{\scriptscriptstyle{K}}}
\newcommand{\LL}{{\scriptscriptstyle{L}}}

\begin{document}
\newcounter{count}

\pagestyle{fancy}\lhead{Orbiting BH binaries and apparent horizons in higher dimensions}
\chead{}
\rhead{\thepage}
\lfoot{}
\cfoot{}
\rfoot{}

\begin{center}
\title{\large Orbiting black-hole binaries and apparent horizons in higher dimensions}
\end{center}

\author{
William G. Cook$^{1}$,
Diandian Wang$^{1}$,
Ulrich Sperhake$^{1,2,3}$
}

\address{$^{1}$~Department of Applied Mathematics and Theoretical Physics,
Centre for Mathematical Sciences, University of Cambridge,
Wilberforce Road, Cambridge CB3 0WA, United Kingdom}

\address{$^{2}$~Theoretical Astrophysics 350-17,
California Institute of Technology,
1200 E California Boulevard, Pasadena, CA 91125, USA}
\address{$^{3}$~Department of Physics and Astronomy, University of Mississippi,
University, Mississippi 38677, USA}

\ead{U.Sperhake@damtp.cam.ac.uk}

\begin{abstract}
We study gravitational wave emission and the structure and formation
of apparent horizons in orbiting black-hole binary systems in
higher-dimensional general relativity. For this purpose we present
an apparent horizon finder for use in higher dimensional numerical
simulations and test the finder's accuracy and consistency in single
and binary black-hole spacetimes. The black-hole binaries we model
in $D=6$ dimensions complete up to about one orbit before merging
or scatter off each other without formation of a common horizon.
In agreement with the absence of stable circular geodesic orbits
around higher-dimensional black holes, we do not find binaries
completing multiple orbits without finetuning of the initial data.
All binaries radiate about $0.13\,\%$ to $0.2\,\%$ of the total
mass-energy in gravitational waves, over an order of magnitude below
the radiated energy measured for four-dimensional binaries. The low
radiative efficiency is accompanied by relatively slow dynamics of
the binaries as expected from the more rapid falloff of the binding
gravitational force in higher dimensions.
\end{abstract}

\maketitle

\section{Introduction}
\label{sec:intro}

Einstein's theory of general
relativity (GR) in four spacetime dimensions is an extraordinarily successful
theory of gravity and
has passed a wealth of tests from solar system dynamics to the bending
of light and the recent detection of gravitational waves (GWs) by
LIGO \cite{Will:2014kxa,Berti:2015itd,Abbott:2016blz,TheLIGOScientific:2016src,TheLIGOScientific:2017qsa}.
While all these tests naturally employ the theory of general relativity
in $D=4$ spacetime dimensions, there is a priori nothing special about
the choice $D=4$.
Rather, $D$ appears merely as a free parameter in the
theory \cite{Emparan:2013moa} which furthermore preserves its fundamental
mathematical properties -- well posedness, diffeomorphism invariance etc. -- for
any value of $D$. In recent decades, higher-dimensional GR with $D>4$
has indeed attracted a lot of attention in many areas of physics research.

Much of the interest in theories with extra dimensions is motivated by
the quest for a unified theory encompassing all physical interactions
and dates back to the pioneering studies of Kaluza and Klein in the
1920s \cite{Kaluza:1921tu,Klein:1926tv}. Extra dimensions naturally
appear in String/M theory \cite{Green:1987sp,Greene:1998ed}
and to some extent in Loop Quantum Gravity \cite{Rovelli:2008zza}.
A particularly fruitful concept arising from these developments is
the {\em holographic principle} according to which the information
of a $D$ dimensional
gravitational system is encoded in its $D-1$ dimensional boundary.
In the form of the Anti-de Sitter/Conformal Field Theory (AdS/CFT)
correspondence
\cite{Maldacena:1997re} and similar dualities, holography
has inspired a great deal of numerical and analytic explorations
of higher dimensional spacetimes containing black holes (BHs) to gain insight
into phenomena such as confinement phase transitions in gauge theories,
non-equilibrium dynamics of quark-gluon plasma or
superconductors; see
e.g.~\cite{Witten:1998zw,Chesler:2008hg,Hartnoll:2009sz}.
Extra dimensions also play a critical role in attempts to
explain the {\em hierarchy problem of physics}, i.e.~the
extraordinary weakness of gravity compared to the other fundamental
interactions of physics. In these scenarios, gravity is diluted at
sub-millimetre scales due to the presence of extra dimensions
such that the fundamental Planck scale is lowered from its effective
four-dimensional value of about $10^{19}~\mathrm{GeV}$ to
the order of Tera electron volts (TeV)
\cite{Antoniadis:1990ew,ArkaniHamed:1998rs,Antoniadis:1998ig,Randall:1999ee,Randall:1999vf}.
In these so-called TeV gravity scenarios, gravity would become the dominant
interaction at energies accessible in particle collisions
at the Large Hadron Collider (LHC) or the interaction of cosmic
rays with the Earth's atmosphere. Despite constraints obtained
at current LHC energies
\cite{Aaboud:2017yvp,Sirunyan:2018xwt},
this leaves open the possibility of
BH production in controlled experiments and, thus, a direct search
for evidence of extra dimensions
\cite{Banks:1999gd,Giddings:2001bu,Dimopoulos:2001hw}.
A quantitative understanding of the dynamics of BH collisions in
higher $D$, particularly the generation of GWs
and the formation of a common BH horizon, is essential in the
interpretation of experimental data \cite{Cardoso:2014uka}
and forms one of the main motivations for our work.
We note in this context that constraints
on extra dimensions have recently been inferred
by Pardo {\em et al.} \cite{Pardo:2018ipy} from the
GW170817 \cite{TheLIGOScientific:2017qsa} observations.
These constraints, however,
do {\em not} hold for extra-dimensional theories with
compact extra dimensions such as the TeV gravity scenarios
mentioned above but rather extend constraints already
known through solar system and laboratory experiments
on macroscopic extra dimensions to astronomical and
cosmological scales; see Sec.~4 in \cite{Pardo:2018ipy} for
details.

Higher dimensional GR has also motivated a large amount of mathematical
studies and revealed important insight into the theory. The space of
solutions to the Einstein equations is exceptionally rich
in $D>4$ compared to the standard four-dimensional case,
including topologically non-spherical BH solutions such
as black rings or black Saturns \cite{Emparan:2001wn,Elvang:2007rd},
but also results in a wider class of instabilities (including
generic violation of cosmic censorship)
limiting the viability of these solutions
\cite{Emparan:2003sy,Lehner:2010pn,Shibata:2010wz,Figueras:2015hkb,Figueras:2017zwa}.
In a complementary approach, Emparan {\em et al.}~~have used the dimensionality
$D$ as an expansion parameter focusing in particular on the large-$D$
limit in the resulting series expansion in $1/D$
\cite{Emparan:2013moa,Emparan:2014cia,Emparan:2015hwa,Emparan:2015gva}.
Aside from probing gravity in generic dimensions, their approach
also recovers rather accurately the physics of GR in $D=4$.
Further details on all these developments in higher-dimensional GR
can be found in the review articles
\cite{Emparan:2008eg,Pretorius:2007nq,Yoshino:2011zza,Sperhake:2013qa,Cardoso:2014uka,Barack:2018yly}.

A key goal for numerical relativity in any given setting, such as higher
dimensions, is to have a full set of diagnostic tools for that
setting. One such tool, and a main focus of this work,
is the ability to find the horizon of a
BH. Once we can find the horizon, we can then
find a measure of the BH's spin and mass, independent of, for instance,
wave extraction calculations. The event horizon of a BH is
a gauge independent object, but one that requires knowledge of
future null infinity to compute; see \cite{Diener:2003jc,Cohen:2008wa}.
From a practical point of view, this involves considerable challenges
in numerical time evolutions, and most applications instead resort to
the computation of apparent horizons (AH);
we follow this approach here.
The AH is a slicing dependent object, that always
lies on or within the event horizon of the BH, defined as
the outermost marginally trapped surface in the spacetime. Developing
efficient, accurate horizon finding has been an important area of
research within numerical relativity, especially in four spacetime
dimensions, with key references including
\cite{Alcubierre:1998rq,Gundlach:1997us,Schnetter:2003pv,Thornburg:1995cp,Thornburg:2003sf};
for a review see \cite{Thornburg:2006zb}. Here we follow the work
of Alcubierre {\em et al.} \cite{Alcubierre:1998rq}, and adapt the algorithm
used for 4D horizon finding to higher dimensions. AH results
have been reported in some higher-dimensional numerical relativity
computations
\cite{Lehner:2010pn,Okawa:2011fv,Figueras:2015hkb,Figueras:2017zwa,Figueras:2017zwa}, but the only detailed description we are aware of is given in the thesis
\cite{Tunyasuvunakool:2017wdi} with a focus on topologically ringlike AHs.
In this paper we therefore
describe in more detail the algorithm used to find AHs in our simulations
of colliding BH binaries in
higher dimensional GR. We apply
the AH finder to
several types of stationary BH data, and time evolutions
of BH binaries and estimate the numerical accuracy of the physical results
extracted from the horizon. The construction of initial data describing
a BH binary with {\em non-zero} orbital angular momentum is a second
main goal of this paper and enables us to compute numerically
the AH of a rotating BH dynamically
formed in higher dimensions in a binary coalescence.
The only simulations of higher-dimensional
BH binary systems {\em with} angular momentum
we are aware of have been reported by Okawa {\em et al} \cite{Okawa:2011fv}.
They simulate grazing collisions of BHs in $D=5$ with initial data
composed of two superposed boosted single-BH spacetimes with
no application of a constraint solving process. Their study
identifies the formation of super-Planckian curvature in a visible domain
in scattering configurations and determines the scattering threshold
for collisions around $\sim 50\,\%$ of the speed of light, but reports
no results on GW emission.
Here, we present
a first exploration of inspiraling BH binary configurations in
$D=6$ and contrast its phenomenology and release of GW
energy with that of four-dimensional inspirals
as observationally confirmed through GW150914 and other events detected by
LIGO/Virgo.

This paper is organised as follows. In Sec.~\ref{sec:notation}, we set the
ground by introducing our notation and, in particular, the index ranges
required in the dimensional reduction in $D>4$ numerical relativity.
Section \ref{sec:HF} contains the theoretical framework of the AH
computations in $D>4$, the particulars encountered in the dimensional
reduction and derives coordinate invariant expressions for extracting
the mass and spin in the limit of stationary BHs. In Sec.~\ref{sec:results}
we test the consistency and accuracy of the AH finder by computing horizons
of stationary single BH spacetimes.
As a further test, we calculate the convergence properties of the
horizon diagnostics in the merger of orbiting binaries in
Sec.~\ref{sec:results2}. There, we also study the dynamics of
binaries with different initial values of the angular momentum,
compute the
GW energy emitted in the evolution
and determine the mass and spin of the merger remnant BHs.
We conclude with a summary of our findings and an outlook for future
work in Sec.~\ref{sec:conclusions}. Details of the construction of
Bowen-York type initial data for BH binaries with non-zero orbital
angular momentum are presented in \ref{sec:BYID}.

\section{Notation}
\label{sec:notation}
Let ($\mathcal{M}, g_{AB}$), $A,\,B,\,\ldots = 0,\,\ldots,\,D-1$, be a $D$ dimensional spacetime, with a Lorentzian metric that solves the $D$-dimensional Einstein equations in vacuum, with vanishing cosmological constant,
\begin{equation}
G_{\A\B}=R_{\A\B}-\frac{1}{2}Rg_{\A \B} = 0\,.
\label{eq:EinsteinD}
\end{equation}
Here we use units where the
gravitational constant and the speed of light $G=c=1$.
Following the standard spacetime decomposition of Arnowitt, Deser and Misner (ADM) \cite{Arnowitt:1962hi}, in the formulation of York \cite{York1979}, we can write the line element as 
\begin{equation}
ds^2 = g_{\A\B}dx^{\A}dx^{\B}
= (-\alpha^2 + \beta_{\I}\beta^{\I})dt^2
+ 2\beta_{\I} dx^{\I} dt + \gamma_{\I\J} dx^{\I}dx^{\J}\,,
\label{eq:ADMlineelement}
\end{equation}
where $I,\,J,\,\ldots = 1,\,\ldots,\,D-1$ and $\alpha$ and $\beta^{\I}$ denote the
lapse function and shift vector respectively, and $\gamma_{\I\J}$
is the induced spatial metric on hypersurfaces given by
$t={\rm const}$. For this choice of coordinates and variables,
the Einstein equations (\ref{eq:EinsteinD}) result in one
{\em Hamiltonian} and $D-1$ {\em momentum} constraints as well
as $D(D-1)/2$ evolution equations cast into first-order-in-time form
by introducing the extrinsic curvature $K_{\I\J}$ through
\begin{equation}
\partial_t \gamma_{\I\J} = \beta^{\M}\partial_{\M} \gamma_{\I\J}
+ \gamma_{\M\J} \partial_{\I} \beta^{\M}
+ \gamma_{\I\M} \partial_{\J} \beta^{\M}
- 2\alpha K_{\I\J}\,.
\label{eq:EK}
\end{equation}
For a detailed review of this decomposition see
\cite{Gourgoulhon:2007ue,Cardoso:2014uka}.

Fully simulating $D-$dimensional spacetimes is a problem that scales
exponentially with the parameter $D$, so we employ a method of
dimensional reduction in order to alleviate this computational cost.
By considering a restricted class of spacetimes with $SO(D-d)$
rotational symmetry we can use the modified cartoon method
\cite{Pretorius:2004jg,Yoshino:2011zz}, to
reduce our $(D-1)-$dimensional spacelike hypersurface to an effective
$d-$dimensional computational domain assuming
$SO(D-d)$ isometry. In this case, the extra dimensions and their impact on
the dynamics is encoded in a few extra variables representing tensor components
in the off-domain directions. Specifically, for the symmetries under
consideration here, we can write the spatial components of a rank-2
tensor $T_{\I\J}$ as
\begin{equation}
  T_{\I\J} = \begin{pmatrix}T_{ij} & 0\\
      0 & T_{ww} \delta_{ab} \end{pmatrix}\,,
\end{equation}
where $i,\,j,\,\ldots = 1,\,\ldots,\,d$, and
$a,\,b,\,\ldots = d+1,\,\ldots,\,D-1$
and $T_{ww}$ denotes one function whose values we only need to
know on the $d$ dimensional computational domain. Tensor
components with upstairs indices are decomposed in the same way
and one straightforwardly shows that $T^{ww}=\gamma^{ww}\gamma^{ww}T_{ww}$,
while the components of vectors in the extra dimensions vanish due
to the symmetry: $V^{\I} = (V^i,~0)$.

In the calculations and simulations presented
in this paper we set $d=3$, restricting our attention to a class
of spacetimes encompassing many of the physical phenomena we wish
to apply our AH finder to, such as BH-BH inspirals,
high energy BH collisions and BHs spinning in a single plane.
We implement the evolution equations
in the form given by Baumgarte, Shapiro, Shibata and Nakamura (BSSN)
\cite{Baumgarte:1998te} \cite{Shibata:1995we}, with the particular
implementation of the BSSN evolution equations in the modified
cartoon formalism used in our code given in \cite{Cook:2016soy}. Henceforth,
we follow the notation in that work which we summarise
as follows.
\begin{itemize}
  \item Upper case early Latin indices $A,\,B,\,C,\,\ldots$ range over the
  full spacetime from 0 to $D-1$.
  \item Upper case middle Latin indices $I,\,J,\,K,\,\ldots$ denote all
  spatial indices, inside and outside the effective three dimensional
  computational domain, running from 1 to $D-1$.
  \item Lower case middle Latin indices $i,\,j,\,k,\,\ldots$ denote indices
  in the spatial computational domain and run from $1$ to $3$, i.e. 
  $x^i=(x,\,y,\,z)$.
  \item Lower case early Latin indices $a,\,b,\,c,\,\ldots$ denote
  spatial indices outside the computational domain, ranging from $4$ to $D-1$,
  i.e. $x^a = (w^4,\ldots,w^{D-1})$.
  \item Greek indices $\alpha,\,\beta,\,\ldots$
  denote all angular directions and range from $2$ to $D-1$.
  \item In our simulations we refer to two coordinate systems, a Cartesian system,
  $X^A = (t,x^i, x^a)$, and a spherical system $Y^A = (t, r, \phi^{\alpha})$.
  \item $\nabla_{\A}$ denotes the covariant derivative in the full
  $D$ dimensional spacetime, whereas $D_{\I}$ denotes the covariant
  derivative on a spatial hypersurface and is calculated from
  the spatial metric $\gamma_{\I\J}$.
\end{itemize}
%


\section{Horizon Finding algorithm}
\label{sec:HF}

\subsection{Horizon finding in higher dimensions}
\label{sec:HFD}

The apparent horizon of a BH is defined as the outermost
marginally trapped surface in the spacetime. Equivalently this is
the surface
on which the expansion of outgoing normal null geodesics is equal
to 0. In order to find the AH on a spacelike hypersurface
$\Sigma$ we calculate the expansion $\Theta = \nabla_{\A} k^{\A}$ of a
congruence of null geodesics with tangent vector $k^{\A}$
moving in the outward normal
direction to a surface $S$, with outward unit normal vector $s^{\A}$.
The calculation of the expansion of this congruence in higher
dimensions proceeds identically to the calculation in 4D, and we present
it here for completeness, following the derivation of Gundlach
\cite{Gundlach:1997us}. Consider a $D$-dimensional spacetime
$(\mathcal{M}, g_{\A\B})$, with covariant derivative $\nabla_{\A}$. We
foliate this spacetime with $D-1$ dimensional spacelike hypersurfaces
$\Sigma_t$ with timelike normal $n^{\A}$. The induced metric on these
hypersurfaces is given by
\begin{equation}
   \gamma_{\A\B} = g_{\A\B}+n_{\A}n_{\B},
\end{equation}
with extrinsic curvature
\begin{equation}
  K_{\A\B} = -\gamma^{\CC}{}_{\A}\nabla_{\CC} n_{\B} = - D_{\A} n_{\B},
\end{equation}
where $D_{\A}$
is the covariant derivative associated to $\gamma_{\A\B}$.  Now let
$S$ be a closed, $D-2$ dimensional, spacelike hypersurface of
$\Sigma_t$, with unit outward spacelike normal $s^{\A}$, which is also
normal to $n^{\A}$. $\gamma_{\A\B}$ induces a metric $q_{\A\B}$ on $S$,
\begin{equation}
  q_{\A\B} = \gamma_{\A\B} - s_{\A}s_{\B}.
\end{equation}
Now let us consider the future pointing, null geodesic congruence,
whose projection onto $\Sigma_t$ is orthogonal to $S$ and $k^{\A}$. $k^{\A}$
satisfies the following equations:
\begin{equation}
  k^{\A}\nabla_{\A}k^{\B} = 0\,,~~~~~k^{\A}k_{\A} = 0\,,~~~~~
q_{\A\B}k^{\A}|_S = 0. \label{eq:kortho}
\end{equation}
In consequence of these conditions, we find that, up to a constant factor
here set to 1 without loss of generality,
\begin{equation}
k^{\A}|_S = s^{\A} + n^{\A}\,.
\end{equation}
Now we can express the expansion $\Theta$ in terms of $(D-1)+1$ quantities,
\begin{eqnarray}
  \Theta &=& g^{\A\B}\nabla_{\A}k_{\B} = (\gamma^{\A\B}
        - n^{\A}n^{\B})\nabla_{\A}k_{\B}\nonumber \\
  &=& \gamma^{\A\B}\nabla_{\A}(s_{\B}+n_{\B})
        - (k^{\A}-s^{\A})(k^{\B}-s^{\B})\nabla_{\A}k_{\B} \nonumber\\
  &=& \gamma^{\A\B}\nabla_{\A}s_{\B} + \gamma^{\A\B}\nabla_{\A}n_{\B}
        - s^{\A}s^{\B}\nabla_{\A}n_{\B}\,.
\end{eqnarray}
In a coordinate basis adapted to the space-time split, we can write this
equation in terms of spatial components,
\begin{equation}
  \Theta = D_{\I}s^{\I} + s^{\I}s^{\J}K_{\I\J} - K\,. \label{eq:expansion}
\end{equation}
The outermost surface upon which $\Theta=0$ everywhere will be our AH.
It will prove convenient to parametrize this surface with a function $F(x^{\I})$, such that our surface is given by the solution to the equation
%
$F(x^{\I}) = 0$
%
and we can write,
\begin{equation}
  s^{\I} = \frac{D^{\I}F}{|DF|}\,,~~~~~|DF|:=\sqrt{D_{\J}F\,D^{\J}F\,}\,,
\end{equation}
and Eq.~(\ref{eq:expansion}) can be reframed as a partial differential
equation to be solved for the scalar $F$.

In order to evaluate Eq.~(\ref{eq:expansion}) in the modified
cartoon formalism, we must distinguish between directions inside and
those pointing off the 3D computational domain.
We can then use
the rotational symmetry in the extra dimensions to simplify tensors
as described in Sec.~\ref{sec:notation},
and furthermore
rewrite derivatives in the extra dimensions
in terms of derivatives in our computational domain; the details
for this procedure are given in
\cite{Cook:2016soy}.
The only terms in Eq.~(\ref{eq:expansion}) that will require such treatment
of extra dimensional components
are $D_{\I}s^{\I}$ and the trace of the extrinsic
curvature. The latter is directly obtained
as $K=K^{\I}{}_{\I}=K^i{}_i+(D-4)\gamma^{ww}K_{ww}$ while we write the former
as
\begin{align}
  D_{\I}s^{\I} &= D_is^i + D_as^a \\
  &= \frac{D_iD^iF}{|DF|} - \frac{(D^iF) (D_jF) D_iD^jF}{|DF|^3}
  + (D-4)\frac{\partial^z F}{|DF|z} + \frac{D-4}{2}\gamma^{ww}\partial_k\gamma_{ww}\frac{\partial^k F}{|DF|} \nonumber\,.
\end{align}
In summary the equation we will use to solve for $F$ is
\begin{eqnarray}
  0=\Theta &=& \frac{D_iD^iF}{|DF|} - \frac{(D^iF)(D_jF)D_iD^jF}{|DF|^3}
  + (D-4)\frac{\partial^z F}{|DF|z}
        \nonumber \\
  && + \frac{1}{2}(D-4)\gamma^{ww}\partial_k\gamma_{ww}\frac{\partial^k F}{|DF|}
  + \frac{K_{ij}\partial^iF \partial^jF}{|DF|^2} - K.\label{eq:expansionF}
\end{eqnarray}
At $z=0$ the term $\partial^zF / z$ appears ill-defined. According
to the regularisation procedures laid out in Appendix B of
\cite{Cook:2016soy}, however, we can substitute in the limit of small $z$
\begin{equation}
\lim_{z\rightarrow 0}\frac{\partial^z F}{z} = \lim_{z\rightarrow 0}
        \partial_z\partial^z F.
\end{equation}

\subsection{Minimisation Algorithm}
\label{sec:AHMin}

In order to numerically solve Eq.~(\ref{eq:expansionF}), we have extended the
minimisation algorithm
provided inside the {\sc Cactus} Computational Toolkit
\cite{Allen:1999,Cactusweb} and described in
\cite{Alcubierre:1998rq,Baumgarte:1996hh} to the case
of $D$ dimensions with $SO(D-3)$ isometry. The first step consists in
reparametrizing the function $F$, restricted to the 3D computational
domain, as
\begin{equation}
F(r,\phi^2,\phi^3) = r - h(\phi^2, \phi^3). \label{eq:Fparam}
\end{equation}
We can then expand $h$ in terms of real spherical harmonics
$Y_{lm}(\phi^2,\phi^3)$.
\begin{equation}
  h(\phi^2,\phi^3) = \sum_l \sum_m \sqrt{4\pi} a_{lm}Y_{lm}(\phi^2, \phi^3). \label{eq:sphharm}
\end{equation}
The iterative search for a solution starts with a spherical
trial function for $h$, from which we
calculate $F$, and so $\Theta$, by Eqs.~(\ref{eq:Fparam}),
(\ref{eq:expansionF}). Next, $\Theta$ is interpolated  onto the points at
which $r=h(\phi^2,\phi^3)$, and used to calculate the surface integral of
$\Theta^2$ over this 2D surface. Powell's minimisation
algorithm \cite{Press1989} then leads to the values $a_{lm}$ for which this
integral is minimised. Once a function $F$ giving a
minimum for $\Theta^2$ is found, we must determine whether this is a local
or global minimum. Following \cite{Alcubierre:1998rq}, this is achieved by
recalculating the candidate function $F$
with higher spatial resolution, and more terms in
the spherical harmonic expansion (\ref{eq:sphharm}). If the value
of the integral of $\Theta^2$ continues to decrease to zero, rather
than reaching some non-zero limiting value, it is interpreted as
a global minimum and the corresponding $F$ defines the AH.
The horizon surface then allows us to calculate
further diagnostic quantities as described in the next section.

\subsection{Black hole diagnostics}
\label{sec:BHdiag}

Once we have found the AH we wish to extract physical
diagnostics of BHs from them. When we consider stationary
BHs, such as those produced by exact initial data (e.g.
Secs.~\ref{sec:Schwarzresults} and \ref{sec:MPresults}), we know
that the world tube of the
AH coincides exactly with the event horizon,
see \cite{Hawking:1973uf} for the proof in $D = 4$, and
\cite{Galloway:2011np} for a discussion of the generalisation of this, and related proofs, to higher dimensions.
For BHs produced as the result of mergers in our simulations
(e.g.~Sec.~\ref{sec:BYresults}) we assume that the spacetime
will, after a long enough period of time, be perturbatively close to a
stationary BH, and that in this case the AH will
closely approximate the spatial cross section of an
event horizon. We therefore base our calculation of BH mass and spin
on the assumption that the spacetime describes a stationary BH.

\subsubsection{Non-spinning Black Holes}

For illustration, we first consider non-rotating BHs in $D$ spacetime
dimensions. These are described by the Tangherlini metric
\cite{Tangherlini:1963bw} given in Schwarzschild coordinates by
\begin{equation}
ds^2 = -\left(1 - \frac{\mu}{\tilde{r}^{D-3}}\right)dt^2 + \left(1 - \frac{\mu}{\tilde{r}^{D-3}}\right)^{-1}d\tilde{r} + \tilde{r}^2d\Omega^2_{D-2}, \label{eq:SchwTang}
\end{equation}
where
\begin{equation}
\mu = \frac{16\pi M}{(D-2) \Omega_{D-2}} \label{eq:ADMmass},
\end{equation}
is the mass parameter. $d\Omega_{n}$ is the line element on the
unit $n$-sphere, parameterised by $n$ angular coordinates, $(\phi^2,
\ldots \phi^{n+1})$,
$\Omega_n$ is the surface area of the
unit $n$-sphere, and $M$ is the ADM mass associated to the
spacetime containing the BH with mass parameter $\mu$.
By considering Eq. (\ref{eq:SchwTang}) we can see that the event
horizon of the BH is given by the surface $\tilde{r}_S^{D-3}
= \mu$. We find the area of this surface to be
\begin{equation}
A_{\mathrm{hor}} = \int_{\mathcal{H}}\sqrt{q} d\phi^2\ldots d\phi^{D-1} = \tilde{r}_S^{D-2}\Omega_{D-2}, \label{eq:areaSchw}
\end{equation}
where $q = \det q_{\I\J}$, and $A_{hor}$ is the area of the AH, as this is a stationary BH.
Combining this expression with Eq.~(\ref{eq:ADMmass}), we find
\begin{equation}
M = \frac{D-2}{16\pi}\Omega_{D-2}^{1/(D-2)}A_{\mathrm{hor}}^{(D-3)/(D-2)}. \label{eq:MADMAH}
\end{equation}

\subsubsection{Spinning black holes}
\label{sec:spinBHs}

The Myers-Perry metric for a singly spinning BH
(the higher-dimensional analogue of the Kerr BH)
is given by \cite{Myers:1986un}
\begin{eqnarray}
ds^2 &=& -dt^2+\frac{\mu}{r^{D-5}\Sigma} (dt-a\sin^2\theta\,d\tilde\phi)^2
+\frac{\Sigma}{\Delta} dr^2 + \Sigma d\theta^2 \nonumber \\
&&+ (r^2+a^2)\sin^2\theta\,d\tilde\phi^2
+r^2\cos^2\theta\,d\Omega_{D-4}^2\,, \nonumber \\
\Sigma &=& r^2+a^2\cos^2\theta\,, \nonumber \\[10pt]
\Delta &=& r^2+a^2-\frac{\mu}{r^{D-5}}\,,
\label{eq:MP}
\end{eqnarray}
where $\mu$ is the mass parameter, and $a$ is the spin parameter.
Unlike in $4D$, where the Kerr BH is
the unique uncharged rotating BH solution, in higher
dimensions other solutions with the same mass and spin, such as
black rings \cite{Emparan:2001wn}, or black Saturns \cite{Elvang:2007rd}
can exist. In the discussion of binary mergers below, we assume that the
end product is a Myers-Perry BH. As we shall see, this expectation
is borne out by the results of the AH finder.
In our notation, the angular coordinates of Eq.~(\ref{eq:MP}) are
$\phi^2=\theta$, $\phi^3=\tilde{\phi}$ and $\phi^4, \ldots \phi^{D-1}$
denote the angular coordinates on the $(D-4)$-sphere in the metric. The
ranges of the angular coordinates are $\theta \in [0,\pi/2]$,
$\tilde{\phi},~\phi^{D-1} \in [0,2\pi]$ and all other angles lie
in the interval $[0,\pi]$.
The location of the horizon is given by the largest root of $\Delta = 0$
\begin{equation}
\frac{\mu}{r_+^{D-5}} = r_+^2+a^2\,,
\label{eq:r+D}
\end{equation}
and, following a brief calculation, the horizon area is, similarly to Eq. (\ref{eq:areaSchw}), given by 
\begin{equation}
A_{\mathrm{hor}} = \int_{\mathcal{H}} \sqrt{q}\, d\theta\, d\tilde\phi\, d\phi^4
        \ldots d \phi^{D-1} = r_+^{D-4}(r_+^2+a^2) \Omega_{D-2}
= r_+\mu \Omega_{D-2}\,. \label{eq:aahD}
\end{equation}
To calculate the spin we will need the equatorial circumference
\begin{equation}
  \ell_e = \int_0^{2\pi} \sqrt{g_{\tilde\phi\tilde\phi}}d\tilde\phi\nonumber
  = 2\pi \frac{r_+^2+a^2}{r_+} =2\pi \frac{\mu}{r_+^{D-4}} \,, \label{eq:leD}
\end{equation}
giving us
\begin{equation}
  r_+^{D-3}=\frac{2\pi}{\Omega_{D-2}} \frac{A_{\rm hor}}{\ell_e}\,,
  ~~~~~
  \mu = \frac{A_{\rm hor}}{r_+ \Omega_{D-2}}\,,~~~~~
  a = \sqrt{\frac{\mu}{r_+^{D-5}}-r_+^2}\,. \label{eq:r+}
\end{equation}
Note that Eq.~(\ref{eq:ADMmass}) holds also for stationary, spinning
BHs. Substituting $\mu$ in that expression in terms of $\Omega_{D-2}$
and $A_{\rm hor}$ and finally setting $\ell_e=2\pi r_S$, $r_+=r_S$
for the non-spinning limit, one indeed recovers Eq.~(\ref{eq:MADMAH}).
In order to obtain a dimensionless quantity for the BH rotation, we follow
Eqs.~(21),~(22) in
\cite{Emparan:2008eg} and define the spin parameter
\begin{equation}
  j = c_J^{1/(D-3)}
        \frac{J}{M^{1/(D-3)}\,M}\,,~~~~~~
  c_J = \frac{\Omega_{D-3}}{2^{D+1}}
        \frac{(D-2)^{D-2}}{(D-3)^{(D-3)/2}}\,,~~~~~~J=\frac{2}{D-2}Ma\,.
  \label{eq:jnorm}
\end{equation}
For $D=5$ this implies $j=a/\sqrt{\mu}$ and, hence, $j=1$ for an extremal
Myers-Perry BH.
\footnote{
For BHs in $D\ge 6$ rotating in a single plane,
there exists no upper limit for the spin magnitude and therefore no
extremal configuration one might naturally wish to identify with $j=1$.
}
In $D=4$ spacetime dimensions, however, Eq.~(\ref{eq:jnorm})
yields $j=a\pi/(2\mu)=a\pi/(4M)$ and one might instead use the more
standard $j\equiv a/M$. Since all BH spacetimes discussed in this work have
$D\ge 5$ spacetime dimensions, we employ (\ref{eq:jnorm}) throughout.

\section{Computing horizons in single-black-hole spacetimes}
\label{sec:results}
The numerical simulations discussed in this and the following section
have been performed with the {\sc Lean}
code \cite{Sperhake:2006cy,Sperhake:2007gu} based
on the {\sc Cactus}
computational toolkit \cite{Allen:1999,Cactusweb} and using
mesh refinement provided by {\sc Carpet}
\cite{Schnetter:2003rb,Carpetweb}.
The {\sc Lean} code was originally developed
for BH simulations in $D=4$ dimensions, using the moving puncture method
\cite{Baker:2005vv,Campanelli:2005dd}, and upgraded
to general $D$ spatial dimensions with $SO(D-3)$ isometry in
\cite{Zilhao:2010sr,Cook:2016soy,Cook:2016qnt}.
Here we use the
modified cartoon implementation originally presented in
\cite{Pretorius:2004jg}; see also \cite{Yoshino:2011zz}.

The first two tests of the AH finder involve analytic initial data 
for spacetimes containing a single BH. We test a Schwarzschild-Tangherlini BH in 5 dimensions
with initial data constructed using isotropic coordinates, and a 5 dimensional singly spinning Myers-Perry
BH with initial data in Kerr-Schild coordinates.
In the first example
we use the horizon mass as a diagnostic for the AH finder, and
in the second we use the horizon mass and spin to
analyse the accuracy of our horizon finder.

\subsection{Isotropic Schwarzschild-Tangherlini} \label{sec:Schwarzresults}

In Schwarzschild coordinates the Schwarzschild-Tangherlini metric
(\ref{eq:SchwTang}) is singular at the event horizon.
These coordinates are not suitable for a numerical
computation of the horizon and we consequently
change to isotropic
coordinates (see e.g.~\cite{Dennison:2010wd}),
\begin{equation}
  \tilde{r} = r\left(1 + \frac{\mu}{4r^{D-3}}\right)^{2/(D-3)}\,,
\end{equation}
which results in the line element
\begin{equation}
  ds^2 = - \left( \frac{4r^{D-3} - \mu}{4r^{D-3} + \mu}\right)^2 dt^2 +
  \left(1+\frac{\mu}{4r^{D-3}}\right)^{4/(D-3)}\left(dx^2+dy^2+dz^2+\sum_a
  dw_a^2\right)\,.
\end{equation}
Here $r^2 = \sum_{\I} (x^{\I})^2$ is the isotropic radius and there is now no coordinate singularity at the horizon.
We perform the ADM spacetime decomposition, picking the isotropic time coordinate as the time coordinate of our foliation, from which we can read off our initial data,
\begin{eqnarray}
  &\alpha = \frac{4r^{D-3} - \mu}{4r^{D-3} + \mu}
        \,,~~~~~~~~~~~~~~~~~~~~~~~~~~
  &\beta^{\I} = 0\,,\\
  &\gamma_{\I\J} = \delta_{\I\J}\left(1+\frac{\mu}{4r^{D-3}}\right)^{4/(D-3)}
        \,,
  &K_{\I\J} = 0.
\end{eqnarray}

We use our horizon finder to calculate the BH mass for a
single isotropic Schwarzschild-Tangherlini BH
for a grid configuration with 2 nested grids with radii
$\{(4,\,2)\times(),\,h\}$, using
the notation of Sec.~II F in \cite{Sperhake:2006cy},
in units of the Schwarzschild radius
$r_S=\mu^{1/(D-3)}$. Such a small grid would be impractical for
time evolutions. Here, however, the grid merely serves as a discretized
subset of the hypersurface of constant time from which the horizon finder
obtains through second-order interpolation the spacetime metric and extrinsic
curvature in its iterative computation of the outermost trapped surface.
The uncertainty in the calculation of the AH is then
dominated
by the error incurred in the interpolation and the second-order
discretization employed inside the AH finder.

In order to quantify the accuracy of the AH finder, we
vary the grid resolution $h$ on the inner refinement level and
correspondingly increase the number of angular points in the AH finder
from 50 to 200.
The results are listed in Table \ref{tab:Schwmass} and enable us to
compute the convergence factor $Q$ given by
\begin{equation}
  Q = \frac{M_{1/8} - M_{1/16}}{M_{1/16} - M_{1/32}}\,,
  \label{eq:convQ}
\end{equation}
where $M_h$ is the value of the horizon mass calculated for a given
resolution $h$. Inserting the values of Table \ref{tab:Schwmass} gives
us $Q=4.99$,
close to the value $Q_2=4$ expected for the second-order discretization
in the AH finder.
\begin{table}[t!]
        \centering
        \begin{tabular}{|c|ccc|}
                \hline
                \hline
                $h/r_S$                         &  1/8          & 1/16      & 1/32 \\
                \hline
                $M_{\mathrm{hor}}/M_{ADM}$&  1.001380     & 1.000237 & 1.000008 \\
                \hline
                $Q$ & $Q_2 = 4.00$ & $Q = 4.99$ & \\
                \hline
                \hline
        \end{tabular}    
        \caption{\footnotesize Measured horizon mass of a
        Schwarzschild-Tangherlini BH at different resolutions.
        The convergence factor $Q$ is computed according to
        Eq.~(\ref{eq:convQ}) and shows good agreement with the
        value $Q_2=4$ expected for second-order convergence.}
        \label{tab:Schwmass}
\end{table}
%

\subsection{5D Myers-Perry in Kerr-Schild coordinates} \label{sec:MPresults}

As in the case of the Schwarzschild BH, the numerical calculation
of the AH of a spinning Myers-Perry BH
requires coordinates
that are not singular at the BH horizon. One such set of
coordinates are Kerr-Schild coordinates, in which the metric is
written in the form
\begin{equation}
  ds^2 = (\eta_{\A\B} + Hl_{\A} l_{\B}) dx^{\A} dx^{\B},
  \label{eq:KerrSchild}
\end{equation}
for appropriate $H$, and null vector $l_{\A}$.
Let us specifically consider a $D=5$ singly spinning Myers-Perry
BH in Cartesian coordinates $(t,x,y,z,w)$. The spin parameter
is $a$ and the spin lies purely in the $x-y$ plane.
Following \cite{Myers:1986un} we can write this metric in Kerr-Schild
form, with the functions in Eq.~(\ref{eq:KerrSchild}) given
by
\begin{equation}
  H = \frac{\mu r^2}{\Pi F}\,,~~~~~\Pi = r^2(r^2+a^2)\,,~~~~~
        F = 1 - \frac{a^2(x^2+y^2)}{(r^2+a^2)^2}\,,
\end{equation}
and
\begin{equation}
  l_{\A} = \left(1,~\frac{rx + ay}{r^2+a^2},\frac{ry - ax}{r^2+a^2},~
        \frac{z}{r},~\frac{w}{r} \right),
        \label{eq:KS3}
\end{equation}
where $r$ is given by the solution to the equation $l_{\A}l^{\A} = 0$,
i.e.~$r^4-r^2(\rho^2-a^2)-a^2(z^2+w^2)=0$, where
$\rho^2=x^2+y^2+z^2+w^2$.
Using the same setup as for the Schwarzschild-Tangherlini BH we foliate the spacetime with slices of constant Kerr-Schild time, $t$, read off the
induced metric from the line element, and calculate the initial extrinsic curvature, which for brevity we do not reproduce here.

\begin{table}[t!]
        \centering
        \begin{tabular}{|c|ccc|ccc|}
                \hline
                \hline
                $a/\sqrt{\mu}$ & & 0.1 & & & 0.9  & \\
                $h/r_{\rm S}$ &  1/16 & 1/32 & 1/64 & 1/32 & 1/48 & 1/64 \\
                \hline
                $M_{\mathrm{hor}}/M$& 1.0005025 & 1.0001200 & 1.0000287 & 1.0012295 & 1.0003776 & 1.0000498 \\
                $j_{\rm hor}$& 0.1007076 & 0.1001868 & 0.1000571 & 0.8979883 & 0.8991569 & 0.8995459 \\
                \hline
                \hline
                $Q$ & $Q_2=4$ & $Q_M=4.02$ & $Q_J=4.19$ &
                $Q_2=2.86$ & $Q_M=2.60$ & $Q_J=3.00$ \\
                \hline
                \hline
        \end{tabular}
        \caption{Horizon mass $M_{\rm hor}$ and the extracted
                 dimensionless spin parameter $j_{\rm hor}
                 =a_{\rm hor}/\sqrt{\mu}$
                 as obtained
                 for a Myers-Perry BH in Kerr-Schild coordinates
                 (\ref{eq:KerrSchild}) in $D=5$
                 with $a/\sqrt{\mu}=0.1$ and $0.9$.
                 The bottom row lists the expected convergence
                 factor $Q_2$ for second-order convergence
                 and the measured convergence factors
                 $Q_f = (f_{h_1}-f_{h_2})/(f_{h_2}-f_{h_3})$
                 for mass ($f=M$) and spin ($f=J$).
                 For the large spin $a/\sqrt{\mu}=0.9$,
                 we require higher grid resolution to find
                 the AH; hence the different range
                 of resolutions used in the convergence analysis.
                }
        \label{tab:MPKSresults}
\end{table}
In Table \ref{tab:MPKSresults} we present the calculated angular
momentum and mass for the Myers-Perry BH for different resolutions
and spin values.
The convergence factors listed in the table are calculated
in analogy to Eq.~(\ref{eq:convQ}) above and yield good agreement
with second-order convergence, but note that we require overall
higher grid resolution to achieve comparable accuracy if the BHs
are spinning.

The computation of an AH is a complex operation which raises
the question of computational efficiency. In our test simulations
we find the AH calculation to be slower
by about an order of magnitude compared with Thornburg's exceptionally
efficient
{\sc AHFinderDirect} \cite{Thornburg:1995cp,Thornburg:2003sf} in $D=4$.
In practice, however, we compute
AHs at regular intervals of the order of $\Delta t\sim 0.1\,r_S$
corresponding to once every 64 or 128 time steps of the innermost
refinement level and we
find an increase by about $1\,\%$ in physical
evolution time relative to the case where we perform
the otherwise identical simulation with the AH finder switched off.

\section{Black-hole binaries}
\label{sec:results2}
We now consider the dynamic formation of a rotating
BH through the coalescence of a BH binary with non-vanishing initial
orbital angular momentum or, equivalently, with non-zero impact parameter $b$.
This study serves two purposes: (i) to test the AH finder in a dynamic
scenario where a rotating BH forms and gradually settles down into
a stationary configuration; (ii) to perform a first exploration of
the dynamics of orbiting BH binary systems in higher dimensions.

Before we quantitatively analyse these configurations, however, we
emphasise a few important points about orbiting binaries in $D>4$
dimensions.
In general, we expect this type of BH collisions to yield similar
regimes of scattering and merging configurations in $D>4$
as known in $D=4$ \cite{Shibata:2008rq,Sperhake:2009jz}: below a
scattering threshold, $b<b_{\rm scat}$,
the binary results in a merger while for $b>b_{\rm scat}$, the
constituents will scatter off to infinity. This behaviour has been
observed for $D=5$ grazing collisions in Ref.~\cite{Okawa:2011fv}.
Even without
numerical simulations, however, we immediately notice two major
differences between inspirals or grazing collisions in $D=4$
as compared with their $D>4$ counterparts. (i) Unlike in $D=4$,
there exist no stable circular orbits around a Myers-Perry
BH in $D>4$ \cite{Cardoso:2008bp}, ruling out, for instance, stable
planetary orbits in a $D>4$ solar system. Viable gravity theories
based on higher-dimensional GR therefore
require some kind of screening mechanism limiting the impact of extra dimensions
to very large or small scales. (ii) The second difference is of quantitative
nature and concerns the relatively weaker gravitational binding force
in binary systems in $D>4$.
For any BH binary whose orbit begins close to an unstable
circular  orbit, this implies correspondingly weaker centrifugal
forces and, thus, that the orbital velocity in the inspiral will
be much slower than in $D=4$.
This is, of course, a special manifestation of the well-known
result that in the large $D$ limit, there is no gravitational force
outside the horizon; cf.~\cite{Emparan:2013moa}. In practice, both
features manifest themselves in the dependence of the binary dynamics
on the initial momentum parameters: (i) we need relatively small initial
momenta lest the binary scatters rather than merges and (ii) without
careful finetuning of the initial momentum, we find it hard to obtain
inspirals completing more than a small fraction of an
orbit prior to a rapid plunge phase.

We note that in \cite{Okawa:2011fv}, grazing BH collisions
have been studied in $5D$. In this work it was noted that no
``zoom-whirl'' orbits were found in $5D$. These orbits have been
identified in numerical studies in $4D$, where inspiraling BHs
whirl around each other for a number of orbits before either
merging or scattering to infinity
\cite{Pretorius:2007jn,Healy:2009zm,Gold:2009hr,Sperhake:2009jz}.
Though we cannot make a statement on the existence of such orbits
generically in higher dimensions without fully exploring the parameter
space of initial momenta and impact parameters, and in particular
investigating high energy grazing collisions, we note that the sharp
transition between scattering orbits and mergers that only involve
a single orbit supports the hypothesis that such zoom-whirl orbits
cannot be formed in higher dimensions.

Bearing in mind these considerations, we numerically model orbiting
binaries and compute the AH of post-merger remnant BHs.
We first need initial data describing a realistic snapshot of a BH binary
in orbit.

\subsection{Numerically constructed Bowen-York like data for BH inspiral in $D>4$} \label{sec:BYresults}
In constructing initial data for spacetimes containing multiple
BHs with linear or angular momentum, we follow the
Bowen-York ansatz \cite{Bowen:1980yu,Cook:2000vr} commonly employed in
$D=4$. Assuming a conformally flat metric,
the constraint equations are numerically solved with a particular
ansatz to give initial data that approximate boosted or spinning
BHs. Following Yoshino {\em et al.} \cite{Yoshino:2006kc} and
Zilh\~ao {\em et al.} \cite{Zilhao:2011yc} we implement initial data for
2 Schwarzschild-Tangherlini BHs (non-spinning) each with
linear momentum. Specifically, we extend the formalism of
\cite{Zilhao:2011yc} to arbitrary directions of the initial
linear momentum vector and, correspondingly, non-zero initial orbital
angular momentum.
We thus construct inspiraling binaries (rather than head-on collisions),
whose coalescence we expect to result in a singly spinning, Myers-Perry BH.

This is achieved by assuming the spatial metric is conformally flat,
and the extrinsic curvature is tracefree,
\begin{eqnarray}
\gamma_{\I\J} &=& \psi^{4/(D-3)}\delta_{\I\J}, \\
K_{\I\J} & = & \psi^{-2} \hat{A}_{\I\J},
\end{eqnarray}
where $\psi$ is a conformal factor, and $\hat{A}_{IJ}$ is the conformally rescaled trace free part of the extrinsic curvature. We then make the ansatz 
\begin{eqnarray}
\hat{A}^{\I\J} = &&\frac{4\pi (D-1)}{(D-2) \Omega_{D-2}}
\frac{1}{r^{D-2}} \\
&&\Big[ n^{\I} P^{\J} + n^{\J} P^{\I}
- n_{\M}P^{\M} \hat{\gamma}^{\I\J}
+ (D-3) n^{\I}n^{\J}P^{\M}n_{\M}\Big]\,, \nonumber
\label{eq:BYA}
\end{eqnarray}
where $P^{\I}$ is the momentum vector of the BH and $n^{\I}$ is
the normal radial vector. Finally we solve for $\psi$ in the
constraint equation by means of an elliptic PDE solver provided by
the {\sc Cactus} thorn {\sc TwoPunctures} \cite{Ansorg:2004ds}. For
further details on this initial data see \ref{sec:BYID}.

\subsection{Apparent horizons of merged black-hole binary}
Based on this initial data construction, we have evolved in $D=6$
dimensions the set of
binary configurations summarised in Table \ref{tab:BHB}.
\begin{table}
\centering
\begin{tabular}{|l|ccc|cccc|}
  \hline
  & $x_0/r_{\rm S}$ & $j_{\rm gl}$ & $h/r_{\rm S}$ &
        $M_{\rm hor}/M$ & $j_{\rm hor}$ &
        $E_{\rm rad}/M$ & $t_{\rm m}/r_{\rm S}$ \\
  \hline
  A1 & 3.185 & 0.1646 & 1/64 & 0.9986 & 0.1597 & $1.969 \times 10^{-3}$ & 137 \\
  A2 & 3.185 & 0.1646 & 1/96 & 0.9984 & 0.1577 & $1.975 \times 10^{-3}$ & 136 \\
  A3 & 3.185 & 0.1646 & 1/128& 0.9984 & 0.1573 & $1.975 \times 10^{-3}$ & 135 \\
  \hline
  B1 & 6.186 & 0.1271 & 1/96 & 1.0014 & 0.1215 & $1.376 \times 10^{-3}$ & 836 \\
  B2 & 6.186 & 0.1362 & 1/64 & 0.9986 & 0.1373 & $1.471 \times 10^{-3}$ &2158 \\
  B3 & 6.186 & 0.1362 & 1/96 & 0.9994 & 0.1356 & $1.549 \times 10^{-3}$ &1738 \\
  B4 & 6.186 & 0.1362 & 1/128& 0.9997 & 0.1352 & $1.558 \times 10^{-3}$ &1612 \\
  B5 & 6.186 & 0.1408 & 1/96 & --     & --     & $5.7 \times 10^{-5}$ & --\\
  \hline
\end{tabular}
\caption{Summary of the BH binary configurations simulated in $D=6$
         dimensions. We characterise a simulation
         by the initial BH location $\pm x_0$,
         the initial angular momentum parameter $j_{\rm gl}$ and the
         resolution $h$ on the innermost refinement level.
         The diagnostic variables are the mass $M_{\rm hor}$
         of the common AH (if one forms), the
         dimensionless spin parameter $j_{\rm hor}$ of the
         post-merger BH, the energy $E_{\rm rad}$ radiated in
         GWs and an estimate $t_{\rm m}$ for the
         time to formation of a common horizon obtained here
         as the retarded time corresponding to the maximum in the energy flux
         $dE_{\rm rad}/dt$. This estimate for $t_{\rm m}$
         agrees within a few $r_{\rm S}$
         with the first time the AH finder reports a common AH.
        }
\label{tab:BHB}
\end{table}
In order to obtain dimensionless numbers, we have normalised the parameters
and results for our binary configurations as follows. Mass and energy
are expressed in units of the ADM mass $M$. Through
Eq.~(\ref{eq:ADMmass}) and the relation $r_{\rm S}=\mu^{1/(D-3)}$,
we obtain the Schwarzschild radius associated with the value of the
ADM mass and we express length and time in units of $r_{\rm S}$. Likewise,
we use the single BH relation (\ref{eq:jnorm}) to associate a {\em global},
dimensionless angular momentum parameter $j_{\rm gl}$ to the spacetime's total
angular momentum $J$. Up to a geometric factor of order unity, $j_{\rm gl}$
measures the total angular momentum per ADM mass raised to the power of
$(D-2)/(D-3)$. In summary, we have
\begin{equation}
  r_{\rm S}^{D-3} \equiv \frac{16\pi M}{(D-2)\Omega_{D-2}}\,,~~~~~
  j_{\rm gl} \equiv c_J^{1/(D-3)} \frac{J}{M^{(D-2)/(D-3)}}\,.
\end{equation}

We first consider the binaries labelled A1 to A3 in Table
\ref{tab:BHB} where two non-spinning, equal-mass BHs start at
positions $x_0/r_S = \pm 3.185$ with opposite linear momentum in the
\begin{figure}
  \centering
  \includegraphics[width=250pt]{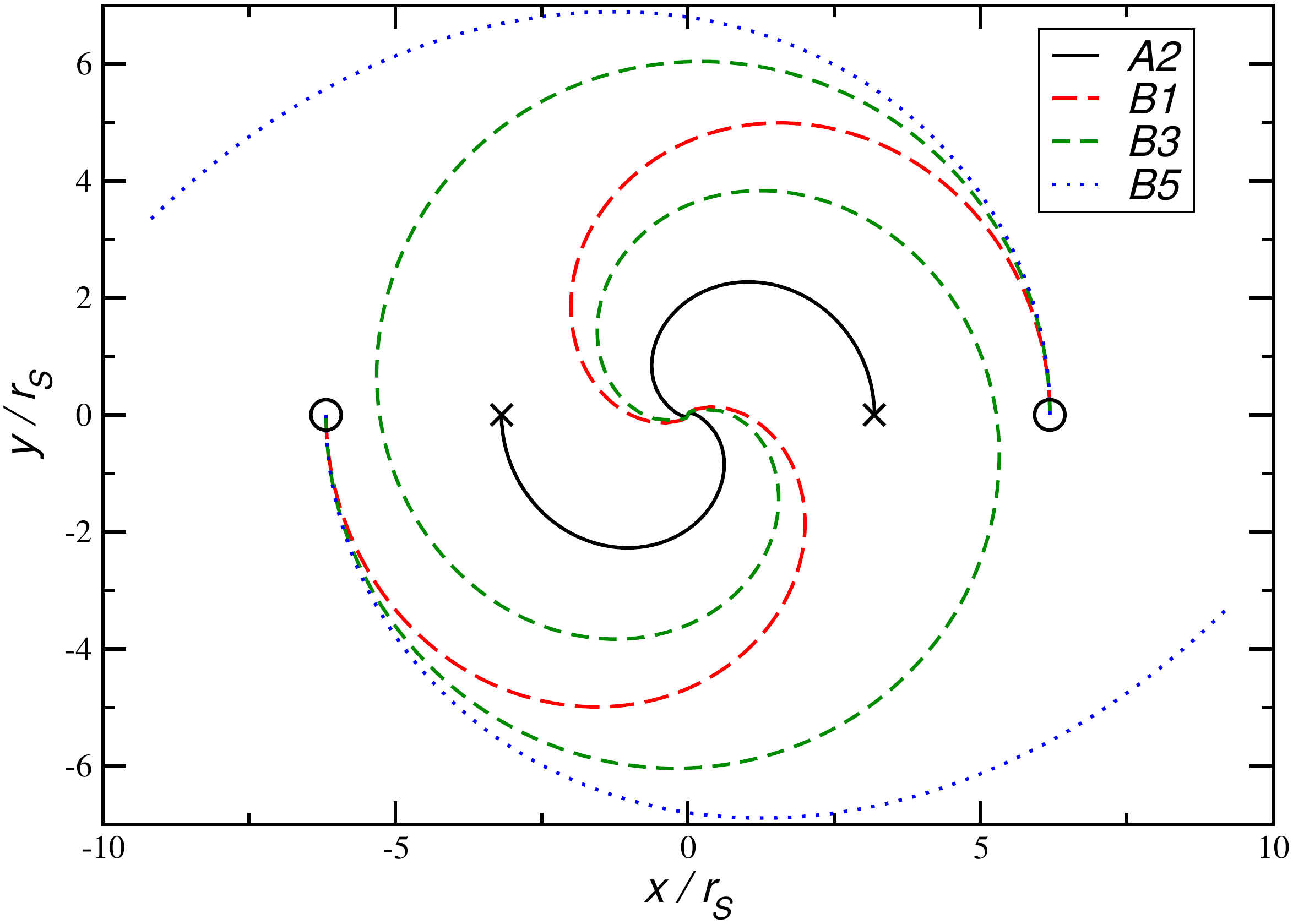}
  \caption{Puncture trajectories of simulations A2, B1, B3 and B5 from
           Table \ref{tab:BHB}. The crosses (circles)
           mark the initial BH positions of configurations A2 (B1,~B3,~B5).
          }
  \label{fig:trajd6}
\end{figure}
$y$ direction corresponding to an angular momentum $j_{\rm gl}=0.1646$.
The grid setup is
$\{(160,\,120,\,72,\,24,\,12,\,6)\times(1.25,\,0.625),~h_i\}$
and the three simulations differ in the
grid resolution: $h_1=r_S/64$,
$h_2=r_S/96$ and $h_3=r_S/128$. The trajectory traced out by the binary
configuration A2 is shown as the solid, black curve in
Fig.~\ref{fig:trajd6}.
Following the procedure described
in \cite{Cook:2016qnt}, we have calculated the energy emitted in
GWs in these simulations and for the post-merger
phase (starting at about $t/r_S = 150$), we extract the spin
of the merger remnant as detailed in Sec.~\ref{sec:spinBHs} above. The results
obtained for the different resolutions are shown in Fig.~\ref{fig:Ej}
together with an analysis of the respective convergence properties.
We find the radiated energy to converge at 4th order and the spin
between 3rd and 4th order which is in agreement with the 4th order
discretization used in the time evolution and indicates that the
error budget here is dominated by the uncertainty of the BH evolution
rather than the AH finder itself. This is confirmed by the
relatively larger uncertainties
of the mass and spin measurements as compared with those obtained for
the single BH cases in Tables \ref{tab:Schwmass} and \ref{tab:MPKSresults}.
\begin{figure}
  \centering
  \includegraphics[width=250pt]{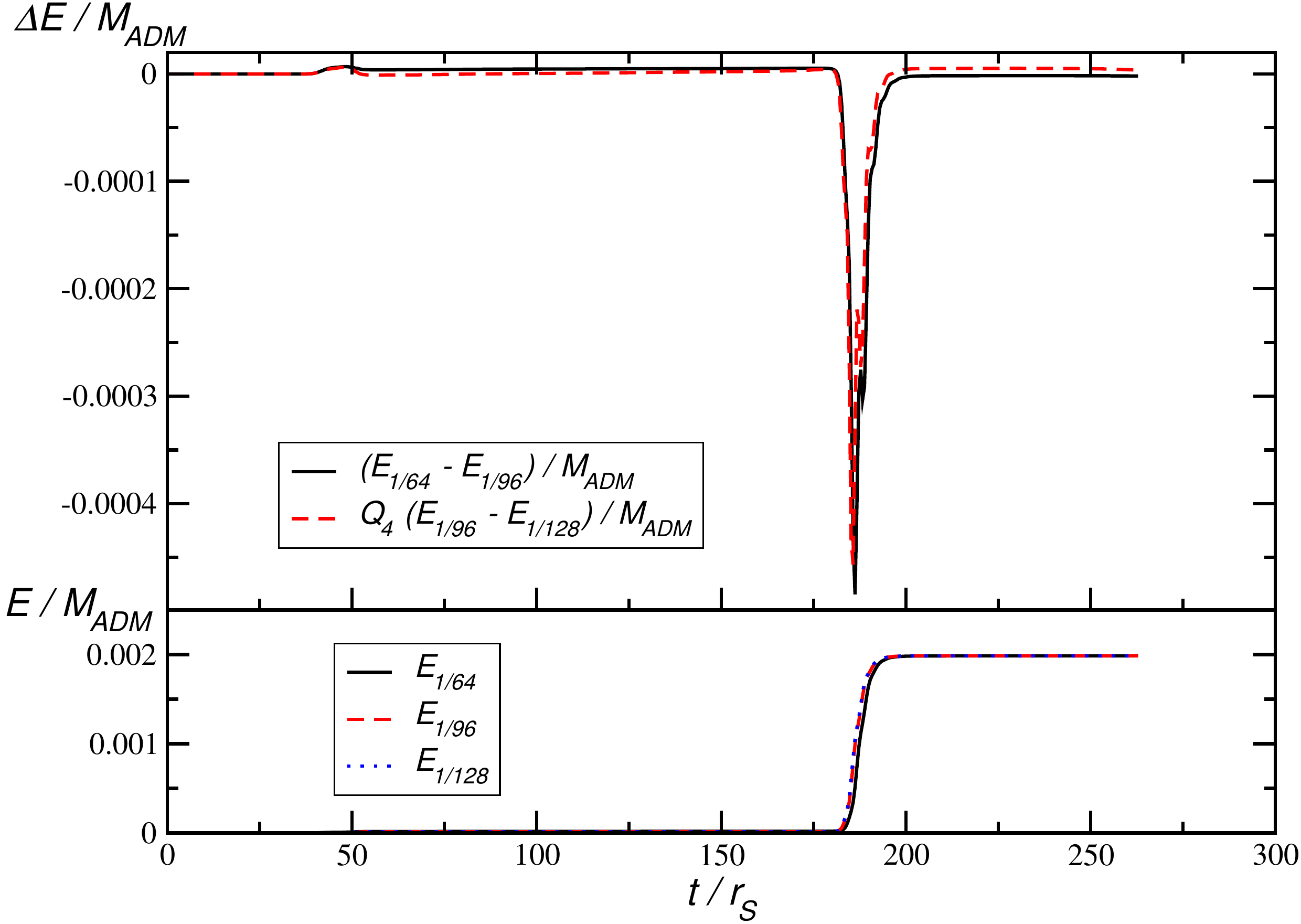} \\[10pt]
  \includegraphics[width=250pt]{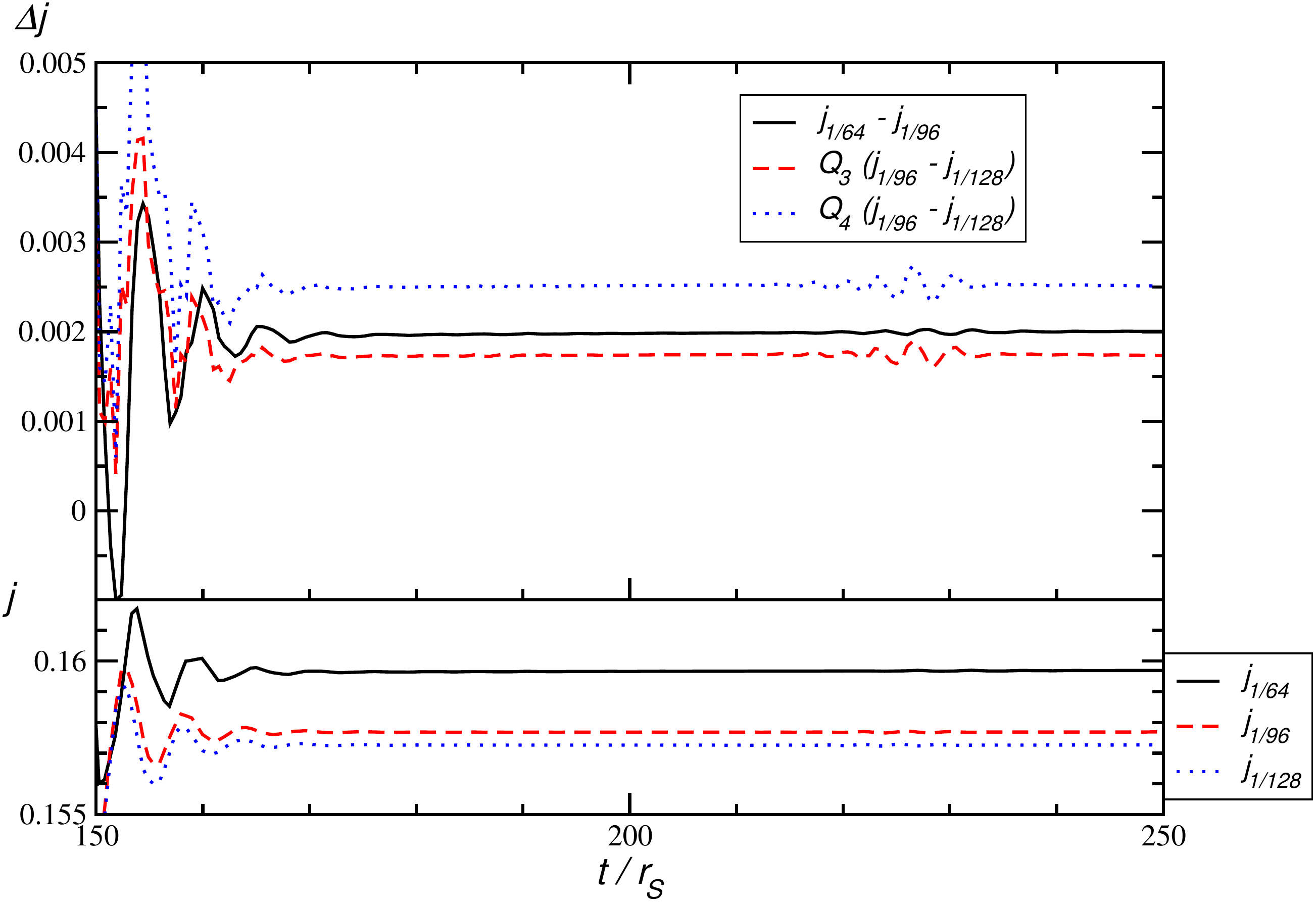}
  \caption{Analysis of the energy radiated in GWs (top)
           and the merger remnant's dimensionless spin (bottom).
           For either quantity, we show results obtained for the three
           grid resolutions in the lower panel and compare in the upper panel
           the differences low-medium vs. medium-high resolution,
           rescaling the latter by a factor $Q_4=5.94$ or
           $Q_3=4.11$ expected for 4th or 3rd order convergence.}
  \label{fig:Ej}
\end{figure}
Comparison with the corresponding Richardson extrapolated values
gives us an uncertainty estimate of $1.8\,\%$ ($0.5\,\%$, $0.22\,\%$)
for the dimensionless spin at low (medium, high) resolution.
We likewise obtain a peak discretization
error of $5\,\%$ ($2\,\%$) for the radiated energy $E_{\rm rad}$
computed for medium (high) resolution, but note that this error
is largely due to the small differences in the time to merger and
the sudden jump of $E_{\rm rad}$ around that time; the uncertainty
in the total radiated energy is smaller, $1\,\%$ or less for all
resolutions used here.
A further source of uncertainty in the radiated energy is
due to finite extraction radii and determined here
as about $2\,\%$ following Ref.~\cite{Sperhake:2011zz}.
Combining these uncertainties, we obtain
for configuration A3
$E_{\rm rad}/M_{\rm ADM}=0.199\pm 0.005\,\%$,
about twice as much as the head-on value
$E_{\rm rad,ho}/M_{\rm ADM}= 0.0819\,\%$ \cite{Witek:2014mha,Cook:2016qnt}.

\subsection{Orbiting BH binaries}
The BH binary discussed in the previous subsection completed less than
half an orbit before coalescing into a single BH. This may in part be due
to the relatively small initial separation and we now consider binaries
starting from larger initial distance given by $\pm x_0/r_{\rm S}= 6.186$.
These models are labelled B1-B5 in Table \ref{tab:BHB}, start
with angular momentum parameters $j_{\rm gl}=0.1271$, $0.1362$ and
$0.1408$, respectively, and have been evolved with a grid setup
$\{(640,\,320,\,160,\,120,\,72,\,24,\,12)\times(2.5,\,1.25,\,0.625),~h_i\}$.
The table lists
the mass and spin parameters $M_{\rm hor}$, $j_{\rm hor}$
obtained from the common AH formed in the coalescence as well as the
radiated energy $E_{\rm rad}$.
We also estimate the time it takes the binaries to merge by
locating the peak value of the radiated energy flux
$dE_{\rm rad}/dt$ in retarded time. We find these values to be
in excellent agreement -- a few $r_S$ -- with the times
the AH finder first finds a common AH. For configuration B5,
we do not find a common horizon which is corroborated by
the binary trajectories shown
in Fig.~\ref{fig:trajd6}.
Although we cannot strictly rule out that binary B5 might merge at a much
later time,
the blue, short-dashed trajectories corresponding to $j_{\rm gl}=0.1408$
indicate a scattering rather than a merging binary. In contrast, binary B1
shows a behaviour similar to that of A2;
the two BHs complete barely more than half an orbit before merging
into a single BH. Increasing the initial angular momentum to the
value $j_{\rm gl}=0.1362$ of case B3, leads to a longer
inspiral phase (cf.~$t_{\rm m}$ in Table \ref{tab:BHB}), but we
still find the binary to merge after only $\sim 1$ orbit.

These findings are compatible with the unstable character of
timelike circular geodesics around BHs in higher dimensions. By finetuning the
angular momentum parameter, it may be possible to obtain binary
systems orbiting multiple times before merger, but small deviations
from such a finetuned value appear to result either in a rapid
plunge (case B3) or a fly-by (case B5). We note in this context that
the time-to-merger $t_{\rm m}$ depends rather sensitively
on the numerical resolution employed (cf.~cases B2 and B4) while
the numerical errors in the other diagnostic quantities remain small.
Close to the threshold that separates mergers
from scattering binaries, even small variations in the angular
momentum (such as those arising from numerical error) can lead
to considerable differences in the trajectories. A more comprehensive
understanding of higher dimensional binaries around the threshold
value of $j_{\rm gl}$ clearly requires a larger number of simulations
and we propose
that the numerical convergence near the threshold then be tested
as in Ref.~\cite{Sperhake:2009jz} through sequences of binaries
rather than one individual configuration.

We finally note the drastically different character of the dynamics we
observe here in $D=6$ dimensions as compared with four
spacetime dimensions. In $D=4$ BH spacetimes have stable
circular orbits and variations in the angular momentum
lead to a continuous transition from inspiraling to plunging
binaries that complete many orbits (depending on initial separation);
see e.g.~Ref.~\cite{Sperhake:2007gu}. The attractive character of
quasi-circular orbits has long since been known in $D=4$ general
relativity as binaries are more efficient in radiating
angular momentum than energy \cite{Peters:1963ux}.
The $D=6$ binaries we have studied here, in contrast, appear
to rapidly scatter off each other or plunge towards merger
instead of approaching a quasi-circular orbit.

As a further consequence of the stronger falloff of gravity in $D>4$,
we note the substantially weaker amount of gravitational radiation
emitted in the $D=6$ binaries studied here:
our merging configurations radiate about
$0.13\,\%$ to $0.2\,\%$ of the total ADM mass,
well below
the $\sim 3\,\%$ found in short inspirals of non-spinning
equal-mass binaries in $D=4$ \cite{Baker:2005vv}. Also,
the energy is radiated entirely in the form of a brief
burst during merger with no significant analogue of the
inspiral contribution clearly perceptible in four-dimensional binaries;
see e.g.~Fig.~18 in Ref.~\cite{Buonanno:2006ui}.
The relatively weaker gravitational
attraction in $D=6$ implies slow orbital motion and, hence,
inefficient generation of GWs except for the
final plunge phase. This is confirmed by the scattering configuration
B5: without the plunge phase, the radiated energy drops by more
than an order of magnitude.

\section{Conclusions}
\label{sec:conclusions}
In this paper we have (i) developed and tested a module for
numerically computing apparent horizons
for topologically spherical BHs in $D>4$ dimensions with $SO(D-3)$
isometry, (ii) generalised the construction of higher-dimensional
Bowen-York type data given in \cite{Zilhao:2011yc} to orbiting BH binaries,
and (iii) used these data and the horizon diagnostics to explore
the dynamics of inspiraling BH binaries in higher dimensional general
relativity. Our main findings are summarised as follows.
\begin{itemize}
  \item The computation of AHs in $D=4$ dimensions based on the
  techniques of Alcubierre \cite{Alcubierre:1998rq} generalise
  straightforwardly to higher-dimensional numerical relativity
  employing the modified cartoon method (Secs.~\ref{sec:HFD},
  \ref{sec:AHMin}).
  \item Mass and spin can be computed directly from the AH's
  surface area and equatorial circumference (Sec.~\ref{sec:BHdiag}).
  \item For analytic
  single BH data, the AH finder obtains the correct values
  with an accuracy of a few times $10^{-5}$ even for modest
  grid resolutions (Tables \ref{tab:Schwmass}, \ref{tab:MPKSresults}).
  In binary evolutions, the error
  budget is dominated by the discretization error of the numerical
  simulation rather than the AH finder itself, leading to
  uncertainties of a few percent in the cases studied here
  (Fig.~\ref{fig:Ej}, Table \ref{tab:BHB}).
  \item The Bowen-York type initial data constructed for axisymmetric
  BH binaries in $D>4$ in Ref.~\cite{Zilhao:2011yc} can be
  generalised straightforwardly to orbiting binaries
  (\ref{sec:BYID}).
  \item In contrast to the $D=4$ dimensional case, we find all
  BH binary systems modelled here in $D=6$ to either merge after
  completing about one orbit or less, or to scatter off each
  other without forming a common horizon.
  Without further finetuning of the initial parameters, we have not
  found binaries completing multiple orbits. This observation
  indicates that the absence of stable circular orbits around
  higher-dimensional BHs generalises to binaries with non-linear
  dynamics (Fig.~\ref{fig:trajd6}).
  \item The orbiting binaries here simulated in $D=6$ spacetime dimensions
  radiate about $0.13\,\%$ to
  $0.2\,\%$ of the ADM energy in GWs,
  about twice as much as in head-on collisions but over an order
  of magnitude less than their $D=4$ counterparts. The energy is
  almost exclusively radiated in the brief plunge-merger phase
  with no analogue to the inspiral contribution present in $D=4$
  (upper panel in Fig.~\ref{fig:Ej}, Table \ref{tab:BHB}).
\end{itemize}
Our numerical study provides a first exploration into the behaviour
of orbiting BH binaries in higher dimensions. Clearly, a much larger set
of runs is required for a comprehensive understanding, especially of the
behaviour near the scattering threshold and the dependency on the
number of spacetime dimensions $D$. Nevertheless, our results
already demonstrate the qualitatively different nature of orbiting
BH binaries in higher dimensions. We opened this work with the observation
that the number $D$ is merely a free parameter in the theory of general
relativity. As far as the dynamics of binary BH spacetimes are concerned,
however, the case $D=4$ appears to be as special as the computation
of geodesics around BHs suggests.

Aside from studying these dynamics in more detail through larger sets of
simulations, our work points to various other extensions. These include
the modelling of spinning binaries,
BH collisions with relatively small impact parameter but much larger initial
boosts, the extraction of angular momentum radiated in GWs
and a multipolar analysis generalising the computation of Kodama-Ishibashi
\cite{Kodama:2003jz} modes from the axisymmetric case of
Ref.~\cite{Witek:2010xi} to the case of orbiting binaries.

\section*{Acknowledgments}
We thank Pau Figueras for the Myers-Perry initial data in Kerr-Schild
coordinates and Nathan Johnson-McDaniel and Markus Kunesch for very
helpful discussions on the topic. We acknowledge financial support
provided under the European Union's H2020 ERC Consolidator Grant
``Matter and strong-field gravity: New frontiers in Einstein's
theory'' grant agreement no. MaGRaTh--646597, funding from the
European Union's Horizon 2020 research and innovation programme
under the Marie Sk\l odowska-Curie grant agreement No 690904, the
COST Action Grant No.~CA16104, from STFC Consolidator Grant No.
ST/L000636/1, the SDSC Comet and TACC Stampede2 clusters through
NSF-XSEDE Award Nos.~PHY-090003,
by PRACE for awarding us access to
MareNostrum at Barcelona Supercomputing Center (BSC), Spain
under Grant No.2016163948, and Cambridge's Cosmos Shared Memory
system through BIS Grant No.~ST/J005673/1 and STFC Grant
Nos.~ST/H008586/1, ST/K00333X/1.
We acknowledge support by the
Yukawa Institute for Theoretical Physics at Kyoto University under
Grant No.~YITP-T-18-05.
W.G.C. acknowledges support by a STFC studentship and D.W. support by
a Trinity College Summer Research Fellowship.

\appendix
\section{Numerically Constructed Initial Data for Non Head-On Collisions} \label{sec:BYID}

In section \ref{sec:BYresults} we have used initial data similar to the well known Bowen-York initial data in 4D to initialise BHs with linear momentum both parallel to and transverse to the direction towards the other BH. The form of this data for higher dimensions was initially proposed in \cite{Yoshino:2006kc,Yoshino2005}, and was explicitly given for momentum only parallel to the direction of the other BH, and implemented in the reduction by isometry dimensional reduction scheme in \cite{Zilhao:2011yc}. Here we explicitly describe the implementation of this data in the modified cartoon formalism, with arbitrary initial momentum in the computational domain.

In order to construct initial data, we must solve the $D$ constraint equations, one Hamiltonian constraint, and $D-1$ momentum constraints, 
\begin{eqnarray}
  \mathcal{H} &=& R + K^2- K^{\I\J}K_{\I\J} = 0\,, \\
  \mathcal{M}^{\I} &=& D_{\J}(K^{\I\J}-\gamma^{\I\J}K) = 0.
\end{eqnarray}
We first decompose the metric in the manner of York and Lichnerowicz, \cite{Lichnerowicz1944,York:1971hw,York:1972sj,York1973}:
\begin{equation}
  \gamma_{\I \J} = \psi^{\frac{4}{D-3}}\hat{\gamma}_{\I\J}\,,~~~~~
  K_{\I\J}=\psi^{-2}\hat{A}_{\I\J}+\frac{1}{D-1}\gamma_{\I\J}
  K\,,
\end{equation}
where $\psi$ is a conformal factor, and $K = \gamma^{\I\J} K_{\I\J}$
is the trace of the extrinsic curvature. We assume that the metric
is conformally flat, $\hat{\gamma}_{\I\J} = \delta_{\I\J}$, and the
maximal slicing condition, that is, $K=0$, which leads to a decoupling of the
Hamiltonian and momentum constraints,
\begin{eqnarray}
  &&\partial_{\I} \hat{A}^{\I\J} = 0, \label{eq:momconst}\\
  && \hat\triangle\psi +
  \frac{D-3}{4(D-2)}\psi^{(-3D-5)/(D-3)}\hat{A}^{\I\J}\hat{A}_{\I\J} =
  0 \label{eq:hamconst}\,.
\end{eqnarray}
Here $\hat\triangle$ is the flat space Laplacian $\hat{\triangle}
\equiv \hat{\gamma}^{\I\J} \partial_{\I} \partial_{\J}$.
Note that the indices on conformally rescaled quantities, such as
$\hat{A}^{\I\J}$ are raised with the conformal metric $\hat{\gamma}^{\I\J}$.
Following Yoshino {\em et al.} \cite{Yoshino:2006kc}, we take an ansatz
for $\hat{A}_{\I\J}$ giving a single boosted BH,
\begin{eqnarray}
  \hat{A}^{\I\J} = &&\frac{4\pi (D-1)}{(D-2) \Omega_{D-2}}
  \frac{1}{r^{D-2}} \\
  &&\Big[ n^{\I} P^{\J} + n^{\J} P^{\I}
  - n_{\M}P^{\M} \hat{\gamma}^{\I\J}
  + (D-3) n^{\I}n^{\J}P^{\M}n_{\M}\Big]\,, \nonumber
\end{eqnarray}
where $P^{\I}$ corresponds to the ADM momentum of the BH and $n^I$ is the normal radial vector in Cartesian coordinates given by
\begin{equation}
n^{\I} \equiv \frac{x^{\I}-x_0^{\I}}{r}\,,~~~~~
r^2 = \delta_{\M\N}x^{\M}x^{\N}\,,
\end{equation}
and $x_0^{\I}$ denotes the position of the BH. 
This ansatz is chosen such that it solves Eq.~(\ref{eq:momconst}), and reproduces the ADM momentum of a boosted BH. We are then left with the task of solving Eq. (\ref{eq:hamconst}), an elliptic PDE for $\psi$. As in 4D for Bowen-York data, we decompose $\psi$ into a Brill-Lindquist component $\psi_{BL}$ \cite{Brill:1963yv}, which on its own gives initial data for a static BH, giving the spacetime approximately the correct ADM mass, and a correction $u$. For a single BH, this is given by
\begin{equation}
  \psi = \psi_{BL} + u = 1 + \frac{\mu}{4r^{D-3}} + u,
\end{equation}
where $\mu$ is the Schwarzschild-Tangherlini mass parameter, and $r$ is the radial distance from the BH. If we wish to solve for more than one BH, clearly Eq. (\ref{eq:momconst}) is linear in $\hat A^{IJ}$, so we can superpose 2 sets of extrinsic curvature
\begin{equation}
\hat A^{\I\J}_{\rm tot}   = \hat{A}^{\I\J}_{(1)} (P^{\K}_{(1)}, n^{\LL}_{(1)} ) + \hat{A}^{\I\J}_{(2)}(P^{\K}_{(2)}, n^{\LL}_{(2)} ),
\end{equation}
where $P_{(i)}^{\K}$ and $n^{\LL}_{(i)}$ are the linear momentum, and radial vector respectively, corresponding to the $i$th BH. We then let $\psi$ take the same form as above, with the Brill-Lindquist term modified to include a contribution from the second BH,
\begin{equation}
  \psi = 1 + \frac{\mu_{(1)}}{4r_{(1)}^{D-3}} + \frac{\mu_{(2)}}{4r_{(2)}^{D-3}} + u,
\end{equation}
where $\mu_{(i)}$ and $r_{(i)}$ are respectively the mass parameter of, and radial distance from, the $i$th BH.
  Now we must solve Eq. (\ref{eq:hamconst}) for $u$, which we achieve using a spectral elliptic PDE solver detailed in  \cite{Ansorg:2004ds,Zilhao:2011yc}, implemented in the {\sc Cactus} thorn {\sc TwoPunctures}.
  
  The final step is to identify how $\hat{A}^{\I\J}$ simplifies within the symmetry restrictions we place on our spacetime in implementing the modified cartoon formalism. Without loss of generality let us consider the case of a single BH with initial momentum and position
\begin{equation}
  P^{\I}=(P_x,\,P_y,\,0,\,0,\,\ldots,\,0)\,,~~~~~
  x_0^{\I}=(x_0,\,0,\,0,\,0,\,\ldots,\,0)\,.
  \label{eq:Px}
\end{equation}
Let us define for convenience
\begin{equation}
  \hat{a}^{\I\J}=n^{\I} P^{\J} + n^{\J} P^{\I}
  - n_{\M}P^{\M} \hat{\gamma}^{\I\J}
  + (D-3) n^{\I}n^{\J}P^{\M}n_{\M}\,,
  \label{eq:BYa}
\end{equation}
so that
\begin{equation}
  \hat{A}^{\I\J} = \frac{4\pi (D-1)}{(D-2)\Omega_{D-2} r^{D-2}}
  \hat{a}^{\I\J} \,. \label{eq:avsA}
\end{equation}
In the modified cartoon approach, we have $w_4=\ldots =w_{D-1}=0$, so
that the radial vector has non-vanishing components only in the $x$,
$y$ and $z$ directions. Furthermore, we use Cartesian coordinates,
so that the expressions we insert into Eq.~(\ref{eq:BYa}) are
given by Eq.~(\ref{eq:Px}) as well as
\begin{eqnarray}
  n^{\I}&=&\left(\frac{x-x_0}{r},~\frac{y}{r},~\frac{z}{r},~0,~\ldots,~0
  \right)\,, \\[10pt]
  \hat{\gamma}_{\I\J}&=&\delta_{\I\J}\,. \label{eq:cfgamma}
\end{eqnarray}

We can now calculate the individual terms in Eq. (\ref{eq:BYa}), firstly for terms inside the computational domain,
\begin{align}
  &n^{i}P^{j} = \left(
  \begin{array}{ccc}
    P_x\frac{x-x_0}{r} &  P_y \frac{x-x_0}{r}& 0 \\[10pt]
    P_x\frac{y}{r}  & P_y \frac{y}{r} & 0 \\[10pt]
    P_x \frac{z}{r} & P_y \frac{z}{r} & 0
  \end{array}
  \right)\,,~~~~~
  n^{j}P^{i} ~=~ \left(
  \begin{array}{ccc}
    P_x \frac{x-x_0}{r} & P_x \frac{y}{r} & P_x \frac{z}{r} \\[10pt]
    P_y \frac{x-x_0}{r} & P_y \frac{y}{r} & P_y \frac{z}{r} \\[10pt]
    0 & 0 & 0
  \end{array}
  \right)\,, \nonumber\\[10pt]
  & n_{\M}P^{\M}\hat{\gamma}^{ij} =
  \left(
  \begin{array}{ccc}
    P_x\frac{x-x_0}{r}+P_y\frac{y}{r} & 0 & 0 \\[10pt]
    0 & P_x\frac{x-x_0}{r}+P_y\frac{y}{r} & 0 \\[10pt]
    0 & 0 & P_x\frac{x-x_0}{r}+P_y\frac{y}{r} \\[10pt]
  \end{array}
  \right)
  \,, \nonumber\\[10pt]
  &(D-3) n^{i}n^{j} n_{\M}P^{\M} = (D-3)\left(P_x\frac{x-x_0}{r}
        +P_y\frac{y}{r}\right)
  \left(
  \begin{array}{ccc}
    \frac{(x-x_0)^2}{r^2} & \frac{(x-x_0)y}{r^2} & \frac{(x-x_0)z}{r^2} \\[10pt]
    \frac{(x-x_0)y}{r^2} & \frac{y^2}{r^2} & \frac{yz}{r^2} \\[10pt]
    \frac{(x-x_0)z}{r^2} & \frac{yz}{r^2} & \frac{z^2}{r^2}
  \end{array}
  \right)\,.
\end{align}
We thus obtain the components
\begin{align}
&\hat{a}^{22} = P_y\frac{y}{r^3}((D-2)y^2+z^2+(x-x_0)^2)+P_x \frac{x-x_0}{r^3}((D-4)y^2-z^2-(x-x_0)^2)\,, \nonumber \\
&\hat{a}^{23} = P_y\frac{z}{r^3}((D-2)y^2+z^2+(x-x_0)^2)+(D-3)P_x\frac{yz}{r^3}(x-x_0)\,, \nonumber \\
&\hat{a}^{12} = P_y\frac{x-x_0}{r^3}((D-2)y^2+z^2+(x-x_0)^2)+P_x\frac{y}{r^3}(y^2+z^2+(D-2)(x-x_0)^2)\,,
\nonumber \\
&\hat{a}^{33} = (P_y\frac{y}{r^3}+P_x\frac{x-x_0}{r^3})(-y^2+(D-4)z^2-(x-x_0)^2)\,,
\nonumber \\
&\hat{a}^{13} = (D-3)P_y\frac{yz}{r^3}(x-x_0)+P_x\frac{z}{r^3}(y^2+z^2+(D-2)(x-x_0)^2)\,,
 \\
&\hat{a}^{11} = P_y\frac{y}{r^3}(-y^2-z^2+(D-4)(x-x_0)^2)+P_x\frac{x-x_0}{r^3}(y^2+z^2+(D-2)(x-x_0)^2)\,.\nonumber
\label{eq:aijgraze}
\end{align}
Finally we calculate the off-domain components,
\begin{equation}
  \hat{a}^{ab}=\left(-P_x\frac{x-x_0}{r}-P_y\frac{y}{r}\right)\delta_{ab}~~~\Rightarrow~~~
\hat{a}^{ww}=-P_x\frac{x-x_0}{r}-P_y\frac{y}{r}\,,
  \label{eq:awwgraze}
\end{equation}
and we note that $\hat{a}^{\I\J}$ is tracefree as expected.

\section*{References}
%

\begin{thebibliography}{100}

\bibitem{Will:2014kxa}
Clifford~M. Will.
\newblock {The Confrontation between General Relativity and Experiment}.
\newblock {\em Living Rev. Rel.}, 17:4, 2014.

\bibitem{Berti:2015itd}
Emanuele Berti et~al.
\newblock {Testing General Relativity with Present and Future Astrophysical
  Observations}.
\newblock {\em Class. Quant. Grav.}, 32:243001, 2015.
\newblock arXiv:1501.07274 [gr-qc].

\bibitem{Abbott:2016blz}
B.~P Abbott et~al.
\newblock {Observation of Gravitational Waves from a Binary Black Hole Merger}.
\newblock {\em Phys. Rev. Lett.}, 116(6):061102, 2016.
\newblock arXiv:1602.03837 [gr-qc].

\bibitem{TheLIGOScientific:2016src}
B.~P. Abbott et~al.
\newblock {Tests of general relativity with GW150914}.
\newblock {\em Phys. Rev. Lett.}, 116(22):221101, 2016.
\newblock arXiv:1602.03841 [gr-qc].

\bibitem{TheLIGOScientific:2017qsa}
B.~P. Abbott et~al.
\newblock {GW170817: Observation of Gravitational Waves from a Binary Neutron
  Star Inspiral}.
\newblock {\em Phys. Rev. Lett.}, 119(16):161101, 2017.

\bibitem{Emparan:2013moa}
R.~Emparan, R.~Suzuki, and K.~Tanabe.
\newblock {The large D limit of General Relativity}.
\newblock {\em JHEP}, 06:009, 2013.
\newblock arXiv:1302.6382 [hep-th].

\bibitem{Kaluza:1921tu}
T.~Kaluza.
\newblock {Zum Unitätsproblem der Physik}.
\newblock {\em Sitzungsber. Preuss. Akad. Wiss. Berlin (Math. Phys.)},
  1921:966--972, 1921.

\bibitem{Klein:1926tv}
O.~Klein.
\newblock {Quantum Theory and Five-Dimensional Theory of Relativity. (In German
  and English)}.
\newblock {\em Z. Phys.}, 37:895--906, 1926.
\newblock [,76(1926)].

\bibitem{Green:1987sp}
M.~B. Green, J.~H. Schwarz, and E.~Witten.
\newblock {\em {Superstring Theory. Vol. 1: Introduction}}.
\newblock Cambridge Monographs on Mathematical Physics. Cambridge University
  Press, Cambridge, UK, 1988.

\bibitem{Greene:1998ed}
B.~R. Greene, D.~R. Morrison, and J.~Polchinski.
\newblock {String theory}.
\newblock {\em Proc. Nat. Acad. Sci.}, 95:11039--11040, 1998.

\bibitem{Rovelli:2008zza}
C.~Rovelli.
\newblock {Loop quantum gravity}.
\newblock {\em Living Rev. Relativity}, 11:5, 2008.

\bibitem{Maldacena:1997re}
J.~M. Maldacena.
\newblock {The large N limit of superconformal field theories and
  supergravity}.
\newblock {\em Adv. Theor. Math. Phys.}, 2:231, 1997.
\newblock hep-th/9711200.

\bibitem{Witten:1998zw}
E.~Witten.
\newblock {Anti-de Sitter space, thermal phase transition, and confinement in
  gauge theories}.
\newblock {\em Adv. Theor. Math. Phys.}, 2:505--532, 1998.
\newblock hep-th/9803131.

\bibitem{Chesler:2008hg}
P.~M. Chesler and L.~G. Yaffe.
\newblock {Horizon formation and far-from-equilibrium isotropization in
  supersymmetric Yang-Mills plasma}.
\newblock {\em Phys. Rev. Lett.}, 102:211601, 2009.
\newblock arXiv:0812.2053 [hep-th].

\bibitem{Hartnoll:2009sz}
S.~A. Hartnoll.
\newblock {Lectures on holographic methods for condensed matter physics}.
\newblock {\em Class. Quant. Grav.}, 26:224002, 2009.
\newblock arXiv:0903.3246 [hep-th].

\bibitem{Antoniadis:1990ew}
Ignatios Antoniadis.
\newblock {A Possible new dimension at a few TeV}.
\newblock {\em Phys. Lett. B}, 246:377--384, 1990.

\bibitem{ArkaniHamed:1998rs}
N.~Arkani-Hamed, S.~Dimopoulos, and G.~R. Dvali.
\newblock {The hierarchy problem and new dimensions at a millimeter}.
\newblock {\em Phys. Lett. B}, 429:263--272, 1998.
\newblock hep-ph/9803315.

\bibitem{Antoniadis:1998ig}
I.~Antoniadis, N.~Arkani-Hamed, S.~Dimopoulos, and G.~R. Dvali.
\newblock {New dimensions at a millimeter to a Fermi and superstrings at a
  TeV}.
\newblock {\em Phys. Lett. B}, 436:257--263, 1998.
\newblock hep-ph/9804398.

\bibitem{Randall:1999ee}
L.~Randall and R.~Sundrum.
\newblock {A large mass hierarchy from a small extra dimension}.
\newblock {\em Phys. Rev. Lett.}, 83:3370--3373, 1999.
\newblock hep-ph/9905221.

\bibitem{Randall:1999vf}
L.~Randall and R.~Sundrum.
\newblock {An alternative to compactification}.
\newblock {\em Phys. Rev. Lett.}, 83:4690--4693, 1999.
\newblock hep-th/9906064.

\bibitem{Aaboud:2017yvp}
Morad Aaboud et~al.
\newblock {Search for new phenomena in dijet events using 37 fb$^{-1}$ of $pp$
  collision data collected at $\sqrt{s}=$13 TeV with the ATLAS detector}.
\newblock {\em Phys. Rev.}, D96(5):052004, 2017.

\bibitem{Sirunyan:2018xwt}
A.~M. Sirunyan et~al.
\newblock {Search for black holes and sphalerons in high-multiplicity final
  states in proton-proton collisions at $\sqrt{s} =$ 13 TeV}.
\newblock 2018.
\newblock arXiv:1805.06013 [hep-ex].

\bibitem{Banks:1999gd}
T.~Banks and W.~Fischler.
\newblock {A Model for High Energy Scattering in Quantum Gravity}.
\newblock 1999.
\newblock hep-th/9906038.

\bibitem{Giddings:2001bu}
S.~B. Giddings and S.~Thomas.
\newblock {High energy colliders as black hole factories: The end of short
  distance physics}.
\newblock {\em Phys. Rev. D}, 65:056010, 2002.
\newblock hep-ph/0106219.

\bibitem{Dimopoulos:2001hw}
S.~Dimopoulos and G.~Landsberg.
\newblock {Black Holes at the LHC}.
\newblock {\em Phys. Rev. Lett.}, 87:161602, 2001.
\newblock hep-th/0106295.

\bibitem{Cardoso:2014uka}
V.~Cardoso, L.~Gualtieri, C.~Herdeiro, and U.~Sperhake.
\newblock {Exploring New Physics Frontiers Through Numerical Relativity}.
\newblock {\em Living Rev. Relativity}, 18:1, 2015.
\newblock arXiv:1409.0014 [gr-qc].

\bibitem{Pardo:2018ipy}
Kris Pardo, Maya Fishbach, Daniel~E. Holz, and David~N. Spergel.
\newblock {Limits on the number of spacetime dimensions from GW170817}.
\newblock {\em JCAP}, 1807(07):048, 2018.

\bibitem{Emparan:2001wn}
R.~Emparan and H.~S. Reall.
\newblock {A Rotating black ring solution in five-dimensions}.
\newblock {\em Phys. Rev. Lett.}, 88:101101, 2002.
\newblock hep-th/0110260.

\bibitem{Elvang:2007rd}
H.~Elvang and P.~Figueras.
\newblock {Black Saturn}.
\newblock {\em JHEP}, 0705:050, 2007.
\newblock hep-th/0701035.

\bibitem{Emparan:2003sy}
R.~Emparan and R.~C. Myers.
\newblock {Instability of ultra-spinning black holes}.
\newblock {\em JHEP}, 0309:025, 2003.
\newblock hep-th/0308056.

\bibitem{Lehner:2010pn}
L.~Lehner and F.~Pretorius.
\newblock {Black Strings, Low Viscosity Fluids, and Violation of Cosmic
  Censorship}.
\newblock {\em Phys. Rev. Lett.}, 105:101102, 2010.
\newblock arXiv:1006.5960 [hep-th].

\bibitem{Shibata:2010wz}
M.~Shibata and H.~Yoshino.
\newblock {Bar-mode instability of rapidly spinning black hole in higher
  dimensions: Numerical simulation in general relativity}.
\newblock {\em Phys. Rev. D}, 81:104035, 2010.
\newblock arXiv:1004.4970 [gr-qc].

\bibitem{Figueras:2015hkb}
P.~Figueras, M.~Kunesch, and S.~Tunyasuvunakool.
\newblock {End Point of Black Ring Instabilities and the Weak Cosmic Censorship
  Conjecture}.
\newblock {\em Phys. Rev. Lett.}, 116(7):071102, 2016.
\newblock arXiv:1512.04532 [hep-th].

\bibitem{Figueras:2017zwa}
P.~Figueras, M.~Kunesch, L.~Lehner, and S.~Tunyasuvunakool.
\newblock {End Point of the Ultraspinning Instability and Violation of Cosmic
  Censorship}.
\newblock {\em Phys. Rev. Lett.}, 118(15):151103, 2017.

\bibitem{Emparan:2014cia}
Roberto Emparan and Kentaro Tanabe.
\newblock {Universal quasinormal modes of large D black holes}.
\newblock {\em Phys. Rev.}, D89(6):064028, 2014.

\bibitem{Emparan:2015hwa}
R.~Emparan, T.~Shiromizu, R.~Suzuki, K.~Tanabe, and T.~Tanaka.
\newblock {Effective theory of Black Holes in the 1/D expansion}.
\newblock {\em JHEP}, 06:159, 2015.
\newblock arXiv:1504.06489 [hep-th].

\bibitem{Emparan:2015gva}
Roberto Emparan, Ryotaku Suzuki, and Kentaro Tanabe.
\newblock {Evolution and End Point of the Black String Instability: Large D
  Solution}.
\newblock {\em Phys. Rev. Lett.}, 115(9):091102, 2015.

\bibitem{Emparan:2008eg}
R.~Emparan and H.~S. Reall.
\newblock {Black Holes in Higher Dimensions}.
\newblock {\em {Living Reviews in Relativity}}, 11(6), 2008.
\newblock {http://www.livingreviews.org/lrr-2008-6}.

\bibitem{Pretorius:2007nq}
Frans Pretorius.
\newblock {Binary Black Hole Coalescence}.
\newblock In M.~{Colpi {\em et al.}}, editor, {\em {Physics of Relativistic
  Objects in Compact Binaries: From Birth to Coalescence}}. Springer, New York,
  2009.
\newblock arXiv:0710.1338 [gr-qc].

\bibitem{Yoshino:2011zza}
H.~Yoshino and M.~Shibata.
\newblock {Exploring Higher-Dimensional Black Holes in Numerical Relativity}.
\newblock {\em Prog.Theor.Phys.Suppl.}, 190:282--303, 2011.

\bibitem{Sperhake:2013qa}
Ulrich Sperhake.
\newblock {Numerical relativity in higher dimensions}.
\newblock {\em Int. J. Mod. Phys. D}, 22:1330005, 2013.
\newblock arXiv:1301.3772 [gr-qc].

\bibitem{Barack:2018yly}
L.~Barack et~al.
\newblock {Black holes, gravitational waves and fundamental physics: a
  roadmap}.
\newblock 2018.
\newblock arXiv:1806.05195 [gr-qc].

\bibitem{Diener:2003jc}
Peter Diener.
\newblock {A New general purpose event horizon finder for 3-D numerical
  space-times}.
\newblock {\em Class. Quant. Grav.}, 20:4901--4918, 2003.
\newblock gr-qc/0305039.

\bibitem{Cohen:2008wa}
M.~I. Cohen, H.~P. Pfeiffer, and M.~A. Scheel.
\newblock {Revisiting Event Horizon Finders}.
\newblock {\em Class. Quant. Grav.}, 26:035005, 2009.
\newblock arXiv:0809.2628 [gr-qc].

\bibitem{Alcubierre:1998rq}
M.~Alcubierre, S.~Brandt, Bernd Br{\"u}gmann, C.~Gundlach, J.~Masso, E.~Seidel,
  and P.~Walker.
\newblock {Test beds and applications for apparent horizon finders in numerical
  relativity}.
\newblock {\em Class. Quant. Grav.}, 17:2159--2190, 2000.
\newblock gr-qc/9809004.

\bibitem{Gundlach:1997us}
C.~Gundlach.
\newblock {Pseudospectral apparent horizon finders: An Efficient new
  algorithm}.
\newblock {\em Phys. Rev. D}, 57:863--875, 1998.
\newblock gr-qc/9707050.

\bibitem{Schnetter:2003pv}
E.~Schnetter.
\newblock {Finding apparent horizons and other two surfaces of constant
  expansion}.
\newblock {\em Class. Quant. Grav.}, 20:4719--4737, 2003.
\newblock gr-qc/0306006.

\bibitem{Thornburg:1995cp}
J.~Thornburg.
\newblock {Finding apparent horizons in numerical relativity}.
\newblock {\em Phys. Rev. D}, 54:4899--4918, 1996.
\newblock gr-qc/9508014.

\bibitem{Thornburg:2003sf}
J.~Thornburg.
\newblock {A Fast apparent horizon finder for three-dimensional Cartesian grids
  in numerical relativity}.
\newblock {\em Class. Quant. Grav.}, 21:743--766, 2004.
\newblock gr-qc/0306056.

\bibitem{Thornburg:2006zb}
J.~Thornburg.
\newblock {Event and Apparent Horizon Finders in 3+1 Numerical Relativity}.
\newblock {\em {Living Reviews in Relativity}}, 10(3), 2007.
\newblock https://doi.org/10.12942/lrr-2007-3.

\bibitem{Okawa:2011fv}
H.~Okawa, K.-i. Nakao, and M.~Shibata.
\newblock {Is super-Planckian physics visible? Scattering of black holes in 5
  dimensions}.
\newblock {\em Phys. Rev. D}, 83:121501, 2011.
\newblock arXiv:1105.3331 [gr-qc].

\bibitem{Tunyasuvunakool:2017wdi}
S.~Tunyasuvunakool.
\newblock {\em {Applications of Numerical Relativity Beyond Astrophysics}}.
\newblock PhD thesis, University of Cambridge, 2017.

\bibitem{Arnowitt:1962hi}
R.~Arnowitt, S.~Deser, and C.~W. Misner.
\newblock {The dynamics of general relativity}.
\newblock In L.~Witten, editor, {\em {Gravitation an introduction to current
  research}}, pages 227--265. John Wiley, New York, 1962.
\newblock gr-qc/0405109.

\bibitem{York1979}
Jr. J.~W. York.
\newblock {Kinematics and dynamics of general relativity}.
\newblock In L.~Smarr, editor, {\em {Sources of {G}ravitational {R}adiation}},
  pages 83--126. Cambridge University Press, Cambridge, 1979.

\bibitem{Gourgoulhon:2007ue}
E.~Gourgoulhon.
\newblock {3+1 {F}ormalism and {B}ases of {N}umerical {R}elativity}.
\newblock 2007.
\newblock gr-qc/0703035.

\bibitem{Pretorius:2004jg}
F.~Pretorius.
\newblock {Numerical relativity using a generalized harmonic decomposition}.
\newblock {\em Class. Quantum Grav.}, 22:425--452, 2005.
\newblock gr-qc/0407110.

\bibitem{Yoshino:2011zz}
H.~Yoshino and M.~Shibata.
\newblock {Higher-Dimensional Numerical Relativity: Current Status}.
\newblock {\em Prog.Theor.Phys.Suppl.}, 189:269--310, 2011.

\bibitem{Baumgarte:1998te}
T.~W. Baumgarte and S.~L. Shapiro.
\newblock {On the {N}umerical integration of {E}instein's field equations}.
\newblock {\em Phys. Rev. D}, 59:024007, 1998.
\newblock gr-qc/9810065.

\bibitem{Shibata:1995we}
M.~Shibata and T.~Nakamura.
\newblock {Evolution of three-dimensional gravitational waves: {H}armonic
  slicing case}.
\newblock {\em Phys. Rev. D}, 52:5428--5444, 1995.

\bibitem{Cook:2016soy}
William~G. Cook, Pau Figueras, Markus Kunesch, Ulrich Sperhake, and Saran
  Tunyasuvunakool.
\newblock {Dimensional reduction in numerical relativity: Modified cartoon
  formalism and regularization}.
\newblock {\em Int. J. Mod. Phys. D}, 25:1641013, 2016.
\newblock arXiv:1603.00362 [gr-qc].

\bibitem{Allen:1999}
{Allen, G. and Goodale, T. and Mass{\'o}, J. and Seidel, E.}
\newblock {The Cactus Computational Toolkit and Using Distributed Computing to
  Collide Neutron Stars}.
\newblock In {\em {Proceedings of Eighth IEEE International Symposium on High
  Performance Distributed Computing, HPDC-8, Redondo Beach, 1999}}, {}, 1999.
  {IEEE Press}.

\bibitem{Cactusweb}
{Cactus Computational Toolkit homepage:} {\tt http://www.cactuscode.org/}.

\bibitem{Baumgarte:1996hh}
T.~W. Baumgarte, G.~B. Cook, M.~A. Scheel, S.~L. Shapiro, and S.~A. Teukolsky.
\newblock {Implementing an apparent-horizon finder in three dimensions}.
\newblock {\em Phys. Rev. D}, 54:4849, 1996.
\newblock gr-qc/9606010.

\bibitem{Press1989}
W.~H. Press, B.~P. Flannery, S.~A. Teukolsky, and W.~T. Vetterling.
\newblock {\em {Numerical {R}ecipes}}.
\newblock Cambridge University Press, Cambridge, 1989.

\bibitem{Hawking:1973uf}
S.~W. Hawking and G.~F.~R. Ellis.
\newblock {\em {The Large Scale Structure of Space-Time}}.
\newblock Cambridge University Press, 1973.

\bibitem{Galloway:2011np}
G.~J. Galloway.
\newblock {Constraints on the topology of higher dimensional black holes}.
\newblock In G.~T. Horowitz, editor, {\em Black holes in higher dimensions},
  pages 159--179, 2012.
\newblock arXiv:1111.5356 [gr-qc].

\bibitem{Tangherlini:1963bw}
F.R. Tangherlini.
\newblock {Schwarzschild field in n dimensions and the dimensionality of space
  problem}.
\newblock {\em Nuovo Cim.}, 27:636--651, 1963.

\bibitem{Myers:1986un}
R.~C. Myers and M.~J. Perry.
\newblock {Black Holes in Higher Dimensional Space-Times}.
\newblock {\em Annals Phys.}, 172:304, 1986.

\bibitem{Sperhake:2006cy}
U.~Sperhake.
\newblock {Binary black-hole evolutions of excision and puncture data}.
\newblock {\em Phys. Rev. D}, 76:104015, 2007.
\newblock gr-qc/0606079.

\bibitem{Sperhake:2007gu}
U.~Sperhake, E.~Berti, V.~Cardoso, J.~A. Gonz{\'a}lez, B.~Br{\"u}gmann, and
  M.~Ansorg.
\newblock {Eccentric binary black-hole mergers: The transition from inspiral to
  plunge in general relativity}.
\newblock {\em Phys. Rev. D}, 78:064069, 2008.
\newblock arXiv:0710.3823 [gr-qc].

\bibitem{Schnetter:2003rb}
E.~Schnetter, S.~H. Hawley, and I.~Hawke.
\newblock {Evolutions in 3-D numerical relativity using fixed mesh refinement}.
\newblock {\em Class. Quant. Grav.}, 21:1465--1488, 2004.
\newblock gr-qc/0310042.

\bibitem{Carpetweb}
{Carpet Code homepage}: {\tt http://www.carpetcode.org/}.

\bibitem{Baker:2005vv}
J.~G. Baker, J.~Centrella, D.-I. Choi, M.~Koppitz, and J.~van Meter.
\newblock {Gravitational-{W}ave {E}xtraction from an inspiraling
  {C}onfiguration of {M}erging {B}lack {H}oles}.
\newblock {\em Phys. Rev. Lett.}, 96:111102, 2006.
\newblock gr-qc/0511103.

\bibitem{Campanelli:2005dd}
M.~Campanelli, C.~O. Lousto, P.~Marronetti, and Y.~Zlochower.
\newblock {Accurate {E}volutions of {O}rbiting {B}lack-{H}ole {B}inaries
  without {E}xcision}.
\newblock {\em Phys. Rev. Lett.}, 96:111101, 2006.
\newblock gr-qc/0511048.

\bibitem{Zilhao:2010sr}
M.~Zilh{\~a}o, H.~Witek, U.~Sperhake, V.~Cardoso, L.~Gualtieri, C.~Herdeiro,
  and A.~Nerozzi.
\newblock {Numerical relativity for D dimensional axially symmetric
  space-times: formalism and code tests}.
\newblock {\em Phys. Rev. D}, 81:084052, 2010.
\newblock arXiv:1001.2302 [gr-qc].

\bibitem{Cook:2016qnt}
William~G. Cook and Ulrich Sperhake.
\newblock {Extraction of gravitational-wave energy in higher dimensional
  numerical relativity using the Weyl tensor}.
\newblock {\em Class. Quant. Grav.}, 34(3):035010, 2017.
\newblock arXiv:1609.01292 [gr-qc].

\bibitem{Dennison:2010wd}
K.~A. Dennison, J.~P. Wendell, T.~W. Baumgarte, and J.~D. Brown.
\newblock {Trumpet slices of the Schwarzschild-Tangherlini spacetime}.
\newblock {\em Phys. Rev. D}, 82:124057, 2010.
\newblock arXiv:1010.5723 [gr-qc].

\bibitem{Shibata:2008rq}
M.~Shibata, H.~Okawa, and T.~Yamamoto.
\newblock {High-velocity collisions of two black holes}.
\newblock {\em Phys. Rev. D}, 78:101501(R), 2008.
\newblock arXiv:0810.4735 [gr-qc].

\bibitem{Sperhake:2009jz}
U.~Sperhake, V.~Cardoso, F.~Pretorius, E.~Berti, T.~Hinderer, and N.~Yunes.
\newblock {Cross section, final spin and zoom-whirl behavior in high-energy
  black hole collisions}.
\newblock {\em Phys. Rev. Lett.}, 103:131102, 2009.
\newblock arXiv:0907.1252 [gr-qc].

\bibitem{Cardoso:2008bp}
V.~Cardoso, A.~S. Miranda, E.~Berti, H.~Witek, and V.~T. Zanchin.
\newblock {Geodesic stability, Lyapunov exponents and quasinormal modes}.
\newblock {\em Phys. Rev. D}, 79:064016, 2009.
\newblock arXiv:0812.1806 [hep-th].

\bibitem{Pretorius:2007jn}
F.~Pretorius and D.~Khurana.
\newblock {Black {H}ole {M}ergers and {U}nstable {C}ircular {O}rbits}.
\newblock {\em Class. Quantum Grav.}, 24:S83--S108, 2007.
\newblock gr-qc/0702084.

\bibitem{Healy:2009zm}
J.~Healy, J.~Levin, and D.~Shoemaker.
\newblock {Zoom-Whirl Orbits in Black Hole Binaries}.
\newblock {\em Phys. Rev. Lett.}, 103:131101, 2009.
\newblock arXiv:0907.0671 [gr-qc].

\bibitem{Gold:2009hr}
R.~Gold and B.~Br{\"u}gmann.
\newblock {Radiation from low-momentum zoom-whirl orbits}.
\newblock {\em Class. Quant. Grav.}, 27:084035, 2010.
\newblock arXiv:0911.3862 [gr-qc].

\bibitem{Bowen:1980yu}
J.~M. Bowen and Jr. J.~W. York.
\newblock {Time-asymmetric initial data for black holes and black-hole
  collisions}.
\newblock {\em Phys. Rev. D}, 21:2047--2056, 1980.

\bibitem{Cook:2000vr}
G.~B. Cook.
\newblock {Initial {D}ata for {N}umerical {R}elativity}.
\newblock {\em {Living Reviews in Relativity}}, 3(5), 2000.
\newblock {http://www.livingreviews.org/lrr-2000-5}.

\bibitem{Yoshino:2006kc}
H.~Yoshino, T.~Shiromizu, and M.~Shibata.
\newblock {Close-slow analysis for head-on collision of two black holes in
  higher dimensions: Bowen-York initial data}.
\newblock {\em Phys. Rev. D}, 74:124022, 2006.
\newblock gr-qc/0610110.

\bibitem{Zilhao:2011yc}
M.~Zilh{\~a}o, M.~Ansorg, V.~Cardoso, L.~Gualtieri, C.~Herdeiro, U.~Sperhake,
  and H.~Witek.
\newblock {Higher-dimensional puncture initial data}.
\newblock {\em Phys. Rev. D}, 84:084039, 2011.
\newblock arXiv:1109.2149 [gr-qc].

\bibitem{Ansorg:2004ds}
M.~Ansorg, B.~Br{\"u}gmann, and W.~Tichy.
\newblock {A single-domain spectral method for black hole puncture data}.
\newblock {\em Phys. Rev. D}, 70:064011, 2004.
\newblock gr-qc/0404056.

\bibitem{Sperhake:2011zz}
U.~Sperhake, B.~Br{\"u}gmann, D.~M{\"u}ller, and C.~F. Sopuerta.
\newblock {11-orbit inspiral of a mass ratio 4:1 black-hole binary}.
\newblock {\em Class. Quant. Grav.}, 28:134004, 2011.
\newblock arXiv:1012.3173 [gr-qc].

\bibitem{Witek:2014mha}
H.~Witek, H.~Okawa, V.~Cardoso, L.~Gualtieri, C.~Herdeiro, M.~Shibata,
  U.~Sperhake, and M.~Zilh{\~a}o.
\newblock {Higher dimensional Numerical Relativity: code comparison}.
\newblock {\em Phys. Rev. D}, 90(8):084014, 2014.
\newblock arXiv:1406.2703 [gr-qc].

\bibitem{Peters:1963ux}
P.~C. Peters and J.~Mathews.
\newblock {Gravitational Radiation from Point Masses in a Keplerian Orbit}.
\newblock {\em Phys. Rev.}, 131:435--439, 1963.

\bibitem{Buonanno:2006ui}
A.~Buonanno, G.~B. Cook, and F.~Pretorius.
\newblock {Inspiral, merger and ring-down of equal-mass black-hole binaries}.
\newblock {\em Phys. Rev. D}, 75:124018, 2007.
\newblock gr-qc/0610122.

\bibitem{Kodama:2003jz}
H.~Kodama and A.~Ishibashi.
\newblock {A master equation for gravitational perturbations of maximally
  symmetric black holes in higher dimensions}.
\newblock {\em Prog. Theor. Phys.}, 110:701--722, 2003.
\newblock hep-th/0305147.

\bibitem{Witek:2010xi}
H.~Witek, M.~Zilh{\~a}o, L.~Gualtieri, V.~Cardoso, C.~Herdeiro, A.~Nerozzi, and
  U.~Sperhake.
\newblock {Numerical relativity for D dimensional space-times: head-on
  collisions of black holes and gravitational wave extraction}.
\newblock {\em Phys. Rev. D}, 82:104014, 2010.
\newblock arXiv:1006.3081 [gr-qc].

\bibitem{Yoshino2005}
H.~Yoshino, T.~Shiromizu, and M.~Shibata.
\newblock {The close limit analysis for head-on collision of two black holes in
  higher dimensions: Brill-Lindquist initial data}.
\newblock {\em Phys. Rev. D}, 72:084020, 2005.
\newblock gr-qc/0508063.

\bibitem{Lichnerowicz1944}
A.~Lichnerowicz.
\newblock {L'integration des {\'e}quations de la gravitation relativiste et le
  probl{\`e}me des $n$ corps}.
\newblock {\em J. Math. Pures et Appl.}, 23:37--63, 1944.

\bibitem{York:1971hw}
Jr. J.~W. York.
\newblock {Gravitational degrees of freedom and the initial-value problem}.
\newblock {\em Phys. Rev. Lett.}, 26:1656--1658, 1971.

\bibitem{York:1972sj}
Jr. J.~W. York.
\newblock {Role of conformal three-geometry in the dynamics of gravitation}.
\newblock {\em Phys. Rev. Lett.}, 28:1082--1085, 1972.

\bibitem{York1973}
Jr. J.~W. York.
\newblock {Conformally invariant orthogonal decomposition of symmetric tensors
  on Riemannian manifolds and the initial-value problem of general relativity}.
\newblock {\em J. Math. Phys.}, 14:456--464, 1973.

\bibitem{Brill:1963yv}
D.~R. Brill and R.~W. Lindquist.
\newblock {Interaction {E}nergy in {G}eometrostatics}.
\newblock {\em Phys. Rev.}, 131:471--476, 1963.

\end{thebibliography}
\bibliographystyle{unsrt}

\end{document}